\documentclass[12pt, draftclsnofoot, onecolumn]{IEEEtran}
\usepackage{amsmath,amsfonts}
\usepackage{amsthm}
\usepackage{algorithmic}
\usepackage{algorithm}
\usepackage{array}
\usepackage[caption=false,font=normalsize,labelfont=sf,textfont=sf]{subfig}
\usepackage{textcomp}
\usepackage{stfloats}
\usepackage{url}
\usepackage{verbatim}
\usepackage{graphicx}
\usepackage{cite}
\usepackage{cuted}
\usepackage{multirow}
\usepackage{enumitem}
\usepackage{xcolor}
\graphicspath{{figs/}}
\newcommand{\tabincell}[2]{\begin{tabular}{@{}#1@{}}#2\end{tabular}}
\newcounter{eqncnttmp}

\begin{document}

\title{How Practical Phase-shift Errors Affect Beamforming of Reconfigurable Intelligent Surface?}

\author{Jun Yang, Yijian Chen, Yijun Cui, Qingqing Wu, Jianwu Dou and Yuxin Wang
\thanks{Jun Yang, Yijian Chen, Yijun Cui, Jianwu Dou and Yuxin Wang are with the State Key Laboratory of Mobile Network and Mobile Multimedia Technology and   the Wireless Product R\&D Institute, ZTE Corporation, Shenzhen, 518055 China (emails: \{yang.jun10, chen.yijian, cui.yijun,dou.jianwu,wang.yuxin\}@zte.com.cn).}
\thanks{Qingqing Wu is with the Department of Electronic Engineering,
Shanghai Jiao Tong University, Shanghai 200240, China (e-mail:
qingqingwu@sjtu.edu.cn).}
\thanks{Manuscript received April 19, 2021; revised August 16, 2021.(Corresponding author: Jun Yang.)}}

\markboth{Journal of \LaTeX\ Class Files,~Vol.~14, No.~8, August~2021}%
{J. Yang \MakeLowercase{\textit{et al.}}: How Practical Phase-shift Errors Affect Beamforming of RIS?}

\IEEEpubid{0000--0000/00\$00.00~\copyright~2021 IEEE}

\maketitle

\begin{abstract}
Reconfigurable intelligent surface (RIS) is a new technique that is able to manipulate the wireless environment smartly and has been exploited for assisting the wireless communications, especially at high frequency band. However, it suffers from hardware impairments (HWIs) in practical designs, which inevitably degrades its performance and thus limits its full potential. To address this practical issue, we first propose a new RIS reflection model involving phase-shift errors, which is then verified by the measurement results from field trials. With this beamforming model, various phase-shift errors caused by different HWIs can be analyzed. The phase-shift errors are classified into three categories: (1) globally independent and identically distributed errors, (2) grouped  independent and identically distributed errors and (3) grouped fixed errors. The impact of typical HWIs, including frequency mismatch, PIN diode failures and panel deformation, on RIS beamforming ability are studied with the theoretical model and are compared with numerical results. The impact of frequency mismatch are discussed separately for narrow-band and wide-band beamforming. Finally, useful insights and guidelines on the RIS design and its deployment are highlighted for practical wireless sytsems.
\end{abstract}


\section{Introduction}\label{sec-intro}
\IEEEPARstart{E}{quiped} with a large number of specially designed passive elements in the size of sub-wavelength, reconfigurable intelligent surface (RIS) has been recognized as a potential key technique for future wireless communications (e.g. \cite{renzo, wuqq0, jianmn, liurq}). By tuning the state of each element, a RIS is able to manipulate the electromagnetic (EM) wave impinging upon it. Different states of the element, controlled by positive intrinsic negative (PIN) diode, varactor diode, liquid crystal, micro-electro-mechanical system (MEMS) or other electronic components (e.g. \cite{zhangl,perez,hum}), correspond to specific changes to the phase, amplitude, polarization or frequency of the incident EM wave. Phase shiftting is one of the most popular manipulation for a reflective or transmissive RIS. With specific phase-shift state arrangement for the element array, the RIS is able to reshape the wave front of the incident EM signal and create an anomalous reflection\cite{diaz} and hence is capable of providing a programmable wireless environment smartly and bring in new communication paradigm\cite{liaskos, renzo2}.

There have been vast studies on system design and optimization for the RIS-assisted wireless communications and perfect hardware condition was usually assumed for the RIS (e.g. \cite{wuqq1,huangcw,houtw}). However, design defect and fabrication errors are often unavoidable, especially for the RIS working at high frequency band. The higher the working frequency, the smaller the element size. For example, a 0.5m $\times$0.5m squared RIS for 30 GHz band consists of about 10,000 elements if the element size is half-wave length ($\approx$ 5 mm). A bias of the metallic patch on an element may be as thin as a single hair and accurate etching or printing is required. In order to balance the fabrication precision and cost, minor errors may be allowed in fabrication stage. On the other hand, RIS panel deformation may occur during the assembling, packaging and transportation. After the RIS installation, thermal expansion and contraction also leads to the accumulation of interior stresses and cause panel deformation. Moreover, the electronic components, such as PIN diodes, embedded in the element may be out of work gradually. Consequently, there are various adverse factors which degrade the performance of a RIS within the life cycle. 

The adverse factors related to the hardware of the MIMO system, such as component non-linearity, I/Q imbalance, quantization error and phase noise, were referred to as hardware impairments (HWIs) and had been recognized as problems that would deteriorate the system performance (e.g. \cite{bjornson, zarei, liuzh, azam} and references therein). As a passive array, RIS has similar HWIs which undoubtedly leads to performance degradation. The HWIs of large intelligent surface (LIS) were modeled as a Gaussian process in \cite{hus, alegria} and the impact was investigated with respect to system capacity and interference. The Gaussian process model was also adopted in \cite{boulogeorgos} for representing the HWIs at transceivers to address the impact on the RIS-assisted system. Secrecy performance analysis for RIS-assisted communications under HWIs of transceivers can also be found in \cite{chenq}. Under the assumption of HWIs, a jointly design of active and passive beamforming for secure RIS-assisted MISO system was proposed in \cite{zhoug}. More discussions on the impact of HWIs on RIS-assisted communication systems can also be found in \cite{liuym,daijx,khalid}. Many of these researches only took into account the HWIs at transceivers while the RIS was assumed in perfect condition. Furthermore, the HWIs in the above works were simply modeled as random noises to the received signal. Such simplified models are far too general and are in lack of quantitative connection to HWIs. How much will the noise level be raised up for a specific HWI (for example, antenna position error in \cite{hus}), or conversely, which error range should the HWIs be confined in in order to guarantee a specific system quality of service?

As to the quantitative analysis of RIS HWIs, the phase-shift error due to quantization is one of the most widely discussed topic. The achievable rate of RIS-assisted single user system under limited phase-shift was discussed in \cite{zhanghl} and they found that the required number of discrete phase-shift for certain data rate was in inverse to the number of RIS elements. A hybrid beamforming scheme was proposed for the RIS-assisted multi-user downlink scenario under the constraint of limited discrete phase shifts in \cite{diby} and the improvement of the sum rate with respect to the number of discrete phase-shifts and RIS size was investigated. Instead of data rate, the diversity of the RIS-assisted communication system was studied in \cite{xup} and the authors concluded that 3 discrete phase-shifts are required to achieve full diversity order. Theoretical beamforming gain loss due to phase quantization was derived in \cite{wuqq2} and optimization algorithm for RIS beamforming was proposed. Other analysis on phase-shift quantization errors can also be found in \cite{lid,gaoh,anjc, sanchez}.

Phase-shift errors due to HWIs rather than quantization were modeled as Von Mises random variables in \cite{badiu} and they found that the transmission through an RIS with a large number of reflecting elements resembled a direct channel with Nakagami scalar fading. Similar phase-shift error model was applied in \cite{wangtx} for system outage probability analysis. The above researches had pointed out that the phase-shift errors will definitely deteriorate the system performance, yet we may still wonder that, behind the performance deterioration, what have happened to the reflection beam under the HWIs and how has the reflection energy been redistributed. Unfortunately, the aforementioned papers do not shed light on the answer. If we have to tolerate the HWIs to some extent, to what extent is the HWIs acceptable for an RIS?

For the RIS with reconfigurable phase-shift state on each element, the HWIs end up acting as phase-shift errors. To the best of our knowledge, the impact of typical HWI on RIS, such as RIS panel deformation and PIN diode failures, have not been reported, yet is closely related to the deployment and maintenance of RIS. On the other hand, the phase-shift errors were modeled as globally independent and identically distributed (i.i.d) variables in most studies. That is, the phase-shift error of each RIS element is represented by the same statistical model. However, different state of the RIS element is fulfilled by distinct patch or circuit configuration and thus a globally identical distribution may not be an appropriate model for the phase-shift error of the RIS element.

Motivated by the above considerations, the impact of practical phase-shift errors caused by various HWIs on the RIS beamforming are investigated in this paper. Unlike some of the previous studies that treated the HWIs as noise terms in channel models, the HWIs are mapped to phase-shift errors in a beamforming model quantitatively. The main contributions of this paper are summarized as follows.
\begin{itemize}
\item 
A RIS beamforming model is proposed and verified with field test. The analytical beamforming gain loss is derived based on this model. Unlike some previous studies, we assume that the elements on a RIS can be divided into different groups regarding the phase-shift error distributions. An inner match factor $\xi$ and a cross match factor $\zeta$ are introduced for the gain loss evaluation. We further prove that the gain loss is independent of the incident angle of the EM wave  for randomly distributed phase-shift errors.
\item 
We sort the practical phase-shift errors due to HWIs into three categories: (1) globally i.i.d errors, (2) grouped i.i.d errors and (3) grouped fixed errors, and we find that the beamforming gain loss due to phase-shift errors of the first category is determined by the inner match factor while the latter two categories rely on both the inner and cross match factors. The analytical expressions of the inner and cross match factors for typical error distributions are also derived for the study of practical cases.
\item 
Practical phase-shift error examples caused by HWIs, including frequency mismatch (both narrow and wide band cases), PIN diode failures and RIS panel deformation are analyzed analytically and numerically. With these analyses, ready-for-use instructions for RIS design, fabrication, deployment and maintenance are provided.
\end{itemize}

The rest of this paper is organized as follows. A RIS beamforming model is established and verified with experimental data in Section \ref{sec-2}. With the beamforming model, theoretical gain loss due to grouped phase-shift errors are derived. In Section \ref{sec-3}, globally i.i.d phase-shift errors, grouped i.i.d phase-shift errors and grouped fixed phase-shift errors are investigated individually and practical phase-shift errors due to HWIs are discussed. Finally, suggestions regarding RIS application in communication systems are concluded in Section \ref{sec-4}.

\emph{Notations}: Scalars are denoted by italic letters, vectors and
matrices are denoted by bold-face lower-case and upper-case
letters, respectively. A uniform distribution in the range $[a,b]$ is represented by $\mathcal{U}(a,b)$ and a normal distribution with mean $\mu$ and standard varience $\sigma$ are denoted by $\mathcal{N}(\mu,\sigma)$. A truncated normal distribution that truncates a normal distribution  $\mathcal{N}(\mu,\sigma)$ in the range $[a,b]$ is denoted by $\mathcal{N}_{t}(\mu,\sigma,a,b)$. The expectation of a random variable $x$ is denoted by $\mathbb{E}(x)$.

\section{Reflection model of RIS}\label{sec-2}
\begin{figure}[h]
\centering{\includegraphics[width=0.9 \textwidth]{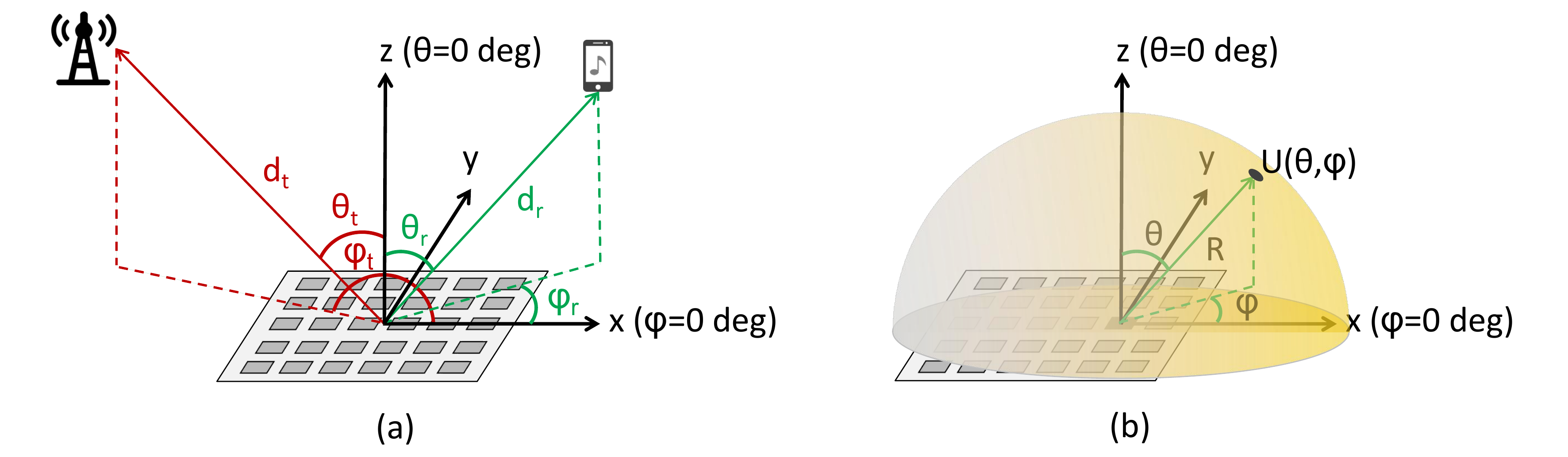}}
\caption{Local coordinates system for (a) an RIS and (b) an RIS element.}
\label{f-coordinates}
\end{figure}
As shown in Fig. \ref{f-coordinates}a, local spherical and Cartesian coordinates with the origin locating at the center of the RIS are adopted in this paper. The location of a certain observation point can be represented by the local spherical coordinates $(d, \theta,\varphi)$, where $d$, $\theta$ and $\varphi$ are the distance from the observation point to the origin, the elevation angle and the azimuth, respectively. The RIS lies in the x-o-y plane, consisting of $N=N_x\times N_y$ elements with element spacing $d_x$ and $d_y$ in x- and y-direction. The locations of the BS and UE are denoted as $\mathbf{r}_{t}=(d_t,\theta_t,\varphi_t)$ and $\mathbf{r}'=(d_r,\theta_r,\varphi_r)$, respectively. When a single element on a RIS is of interest, a local coordinates system as shown in Fig. \ref{f-coordinates}b is applied. The incident or reflection angle at an RIS element is determined by the elevation angle and azimuth $(\theta, \varphi)$ in the local coordinates system.
\subsection{Beamforming Model}
For a reflective RIS, the reflected energy is confined to the front hemisphere towards the positive z-axis , i.e., $\theta\in[0,\frac{\pi}{2}]$, as shown in Fig. \ref{f-coordinates}b. Assuming that the total power of the EM wave reflected by a single RIS element is $P$, then the total power holds when the EM wave propagates to a hemispherical surface with radius $R$, according to the law of energy conservation, which reads
\begin{equation}
P=R^{2}\int_{0}^{2\pi}\int_{0}^{\frac{\pi}{2}}U(\theta,\varphi)\sin\theta d\theta d\varphi,\label{eq-p-reflected}
\end{equation}
where $U(\theta,\varphi)$ is the power density on the hemispherical surface. The power density can be further written as
\begin{equation}
U(\theta,\varphi)=U_mF(\theta,\varphi),\label{eq-p-density}
\end{equation}
 where $U_m$ denotes the maximum power density on the surface whereas $F(\theta,\varphi)$ depicts the reflection pattern of the RIS element. According to \eqref{eq-p-reflected} and \eqref{eq-p-density}, $U_m$ can be determined and the power density can be expressed as
\begin{equation}
U(\theta,\varphi)=\frac{PF(\theta,\varphi)}{R^2\Omega_A}=\frac{PF_{n}(\theta,\varphi)}{R^2},\label{eq-p-density2}
\end{equation}
where $\Omega_A$ is the beam solid angle as defined in \cite{stutzman},
\begin{equation}
\Omega_A=\int_{0}^{2\pi}\int_{0}^{\frac{\pi}{2}}F(\theta,\varphi)\sin\theta d\theta d\varphi,
\end{equation}
and $F_{n}(\theta,\varphi)=F(\theta,\varphi)/\Omega_A$ is the power-normalized pattern.
For an isotropic RIS element, i.e., $F(\theta,\varphi)=1$, the incident EM wave is reflected uniformly towards all directions and the power density is a constant on the hemispherical surface, $U(\theta,\varphi)=U_{m}=P/(2\pi R^2)$. 

As a passive node in the communication system, the RIS does not transmit signal itself but reflects the EM wave impinge upon it from the BS. For a BS locating at $\mathbf{r}_t$ and a RIS element locating at $\mathbf{r}$, if the BS has a single antenna whose power-normalized pattern is $F_{\text{n,tx}}(\theta,\varphi)$, the power received by the RIS element $P_r$ is related to the transmitted power $P_t$ at the BS as
\begin{equation}
P_{r}= S_{e}\frac{P_{t}F_{\text{n,tx}}(\theta^{\text{tx}},\varphi^{\text{tx}})F(\theta^{\text{inc}},\varphi^{\text{inc}})}{|\mathbf{r}-\mathbf{r}_t|^2},\label{eq-p-inc}
\end{equation}
where $(\theta^{\text{tx}},\varphi^{\text{tx}})$ and $(\theta^{\text{inc}},\varphi^{\text{inc}})$ represent the angle of departure at the BS and the angle of arrival at the RIS element and $S_e=dxdy$ is the area of the RIS element. Additionally, the phase delay of the EM wave from the BS to  the RIS element can be calculated as $\phi^{\text{inc}}=k|\mathbf{r}-\mathbf{r}_t|$, where $k=2\pi/\lambda$ is the wave number, $\lambda$ being the wave length.  Finally, the complex amplitude of the EM signal reflected by the $\ell$-th element on the RIS at an observation point $\mathbf{r}'$ can be written as
\begin{equation}
A_{\ell}(\mathbf{r'})=\frac{\sqrt{P_{r,\ell}F_n(\theta_{\ell},\varphi_{\ell})}}{|\mathbf{r}_{\ell}-\mathbf{r}'|}\Gamma_{\ell} e^{j(\phi^{\text{inc}}_{\ell}+k|\mathbf{r}_{\ell}-\mathbf{r}'|)},\label{eq-amp}
\end{equation}
where $\Gamma_{\ell}=|\Gamma_{\ell}| e^{j\phi_{\ell}}$ is the complex reflection coefficient of the RIS element , $\phi_{\ell}$ being the phase-shift imposed by the RIS element to the EM signal. The reflection angle $(\theta_{\ell},\varphi_{\ell})$ is determined by the spatial vector $(\mathbf{r}_{\ell}-\mathbf{r}')$ in the local coordinates of the RIS.

In most cases, the RIS is utilized to enhance signal coverage or tackle the blockage and it is usually deployed far from the BS, hence the BS can be regarded as a point source and \eqref{eq-amp} can be applied. Note that the signal amplitude $A_{\ell}$ in \eqref{eq-amp} is not the exact amplitude of the electric or magnetic field and the transform between the signal amplitude and electric field amplitude can be found in \cite{tangwk}. If a UE equipped with an isotropic antenna is placed at the observation point, the received power at the UE reflected by a RIS with $N$ elements can then be expressed as
\begin{equation}
P(\mathbf{r}')=S\Big|\sum_{\ell=1}^{N}A_{\ell}(\mathbf{r}')\Big|^{2}
=S\Big|\sum_{\ell=1}^{N}\frac{\sqrt{P_{r,\ell}F_n(\theta_{\ell},\varphi_{\ell})}}{|\mathbf{r}_{\ell}-\mathbf{r}'|}\Gamma_{\ell}  e^{j(\phi^{\text{inc}}_{\ell}+k|\mathbf{r}_{\ell}-\mathbf{r}'|+\phi_{\ell})}\Big|^{2},
\label{eq-p-nf}
\end{equation}
where $S$ is the effective aperture of the UE antenna. Obviously, the maximum received power can be obtained if $\phi^{\text{inc}}_{\ell}+k|\mathbf{r}_{\ell}-\mathbf{r}'|+\phi_{\ell}=\phi_{c}$ for $\ell=1, 2, \cdots, N$, where $\phi_{c}$ is a constant phase which can be simply set to be zero. The optimal phase-shift for the $\ell$-th RIS element can then be written as
\begin{equation}
\phi_{\ell}^{*}=-\phi_{\ell}^{inc}-k|\mathbf{r}_{\ell}-\mathbf{r}'|+\phi_c.\label{eq-phi-opt}
\end{equation}
Equation \eqref{eq-phi-opt} is a general near-field solution, but when the receiver locates in the far-field zone of an antenna array, i.e. $|\mathbf{r}'|>2D^2/\lambda$, where D is the largest dimension of the array, a plane wave approximation can be made. Let $(\theta_i,\varphi_i)$ and $(\theta_r,\varphi_r)$ be the incident and reflection angle, respectively, and assume that the RIS element is numbered starting from the lower left corner to the upper right corner in Fig. \ref{f-coordinates}a, then the optimal phase-shift for the element at the n-th row and m-th column is
\begin{equation}
\phi_{nm}^{*}=-k[(N_x-m)c_{x}+(N_y-n)c_{y}],\label{eq-phi-opt-ff}
\end{equation}
where $c_{x}=d_{x}(\sin\theta_i\cos\varphi_i+\sin\theta_r\cos\varphi_r)$, $c_{y}=d_{y}(\sin\theta_i\sin\varphi_i+\sin\theta_r\sin\varphi_r)$.

Ideally, if the phase-shift of each element on the RIS is tuned according to \eqref{eq-phi-opt} to form a precoding matrix, a perfect beamforming is achieved. However, due to the limitation of the hardware, the phase-shift at each element may not be switched to the optimal one. Instead,  a set of discrete phase-shifts is available and the one that is most close to the optimal phase-shift is chosen for the beamforming.

With the beamforming model \eqref{eq-p-nf}, the distribution of the reflected power on a given hemispherical surface corresponding to certain precoding matrix can be calculated. When the distribution of the reflected power on the hemispherical surface is obtain, it forms a 3D beam pattern which shows the exact direction of the reflection beam. Sometimes, the width and gain of the reflection beam is of concern, then we will only gather the reflected power along one of the orthodrome on the hemispherical surface to form a 2D beam pattern. The 2D and 3D beam patterns are referred to as numerical results in this paper and will be applied for phase-shift error analysis in the sequel of this paper. 

\begin{figure}[h]
\centering
\subfloat[]{
\includegraphics[width=0.4 \textwidth]{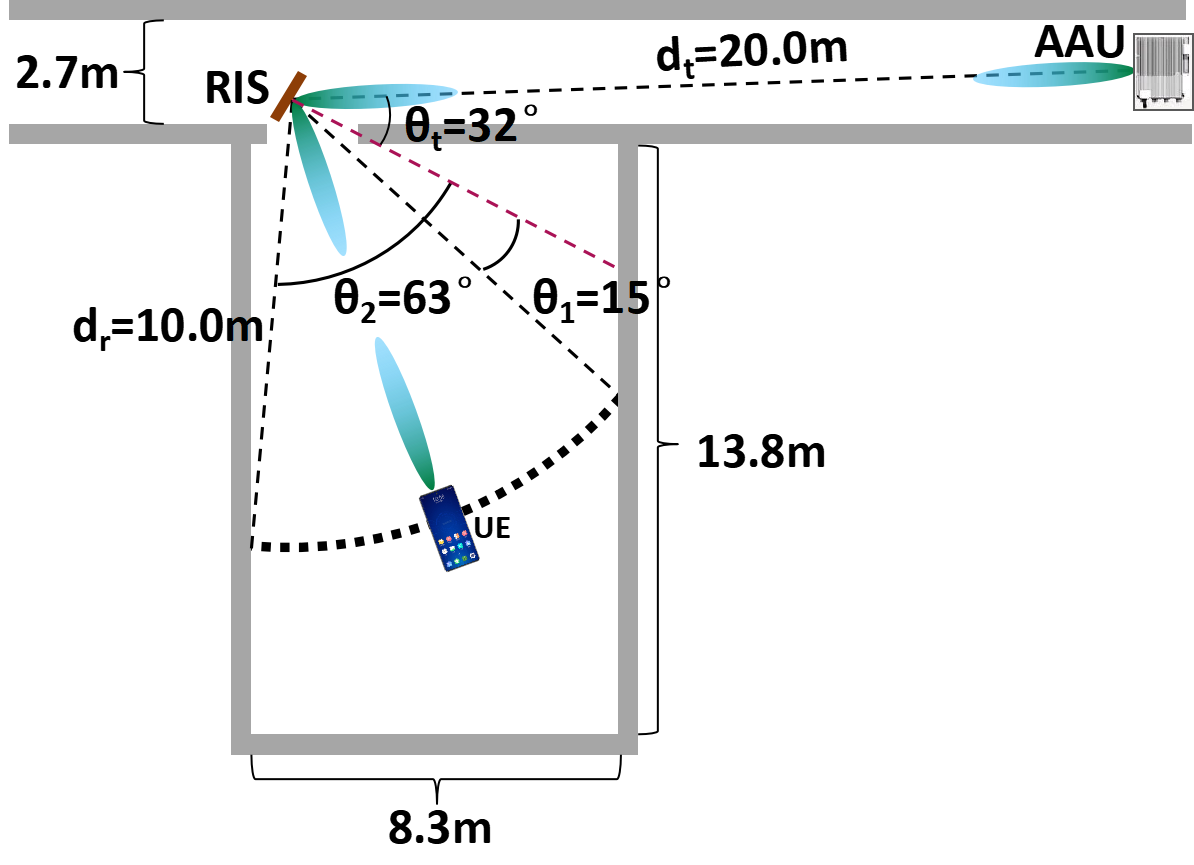}}\;
\subfloat[]{
\includegraphics[width=0.4 \textwidth]{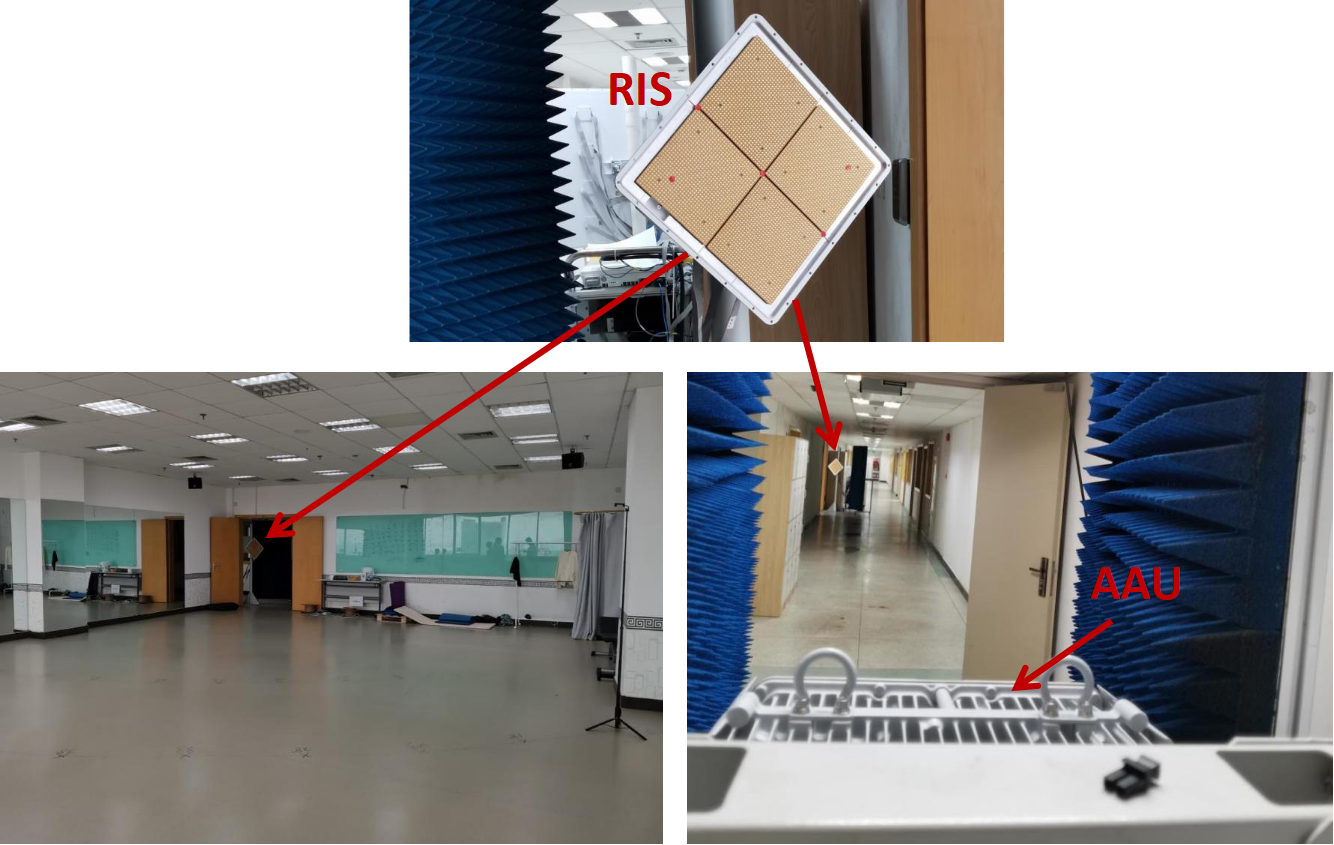}}\;
\subfloat[]{
\includegraphics[width=0.8 \textwidth]{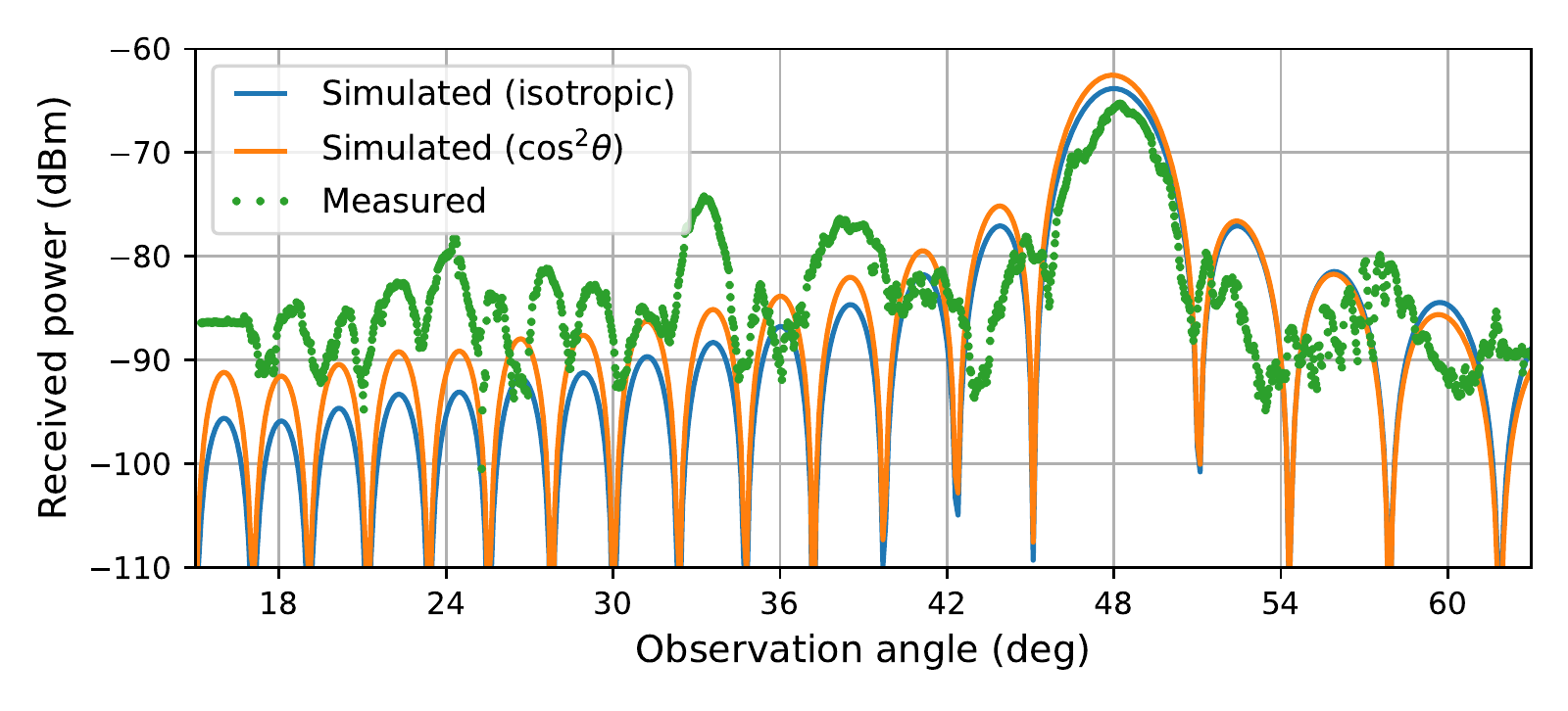}}
\caption{(a) Illustration of the frequency mismatch in wideband system; (b) beamforming loss at different frequency when a precoding matrix designed for the central frequency is applied and (c) measured RSRP versus simulation result.}
\label{f-test}
\end{figure}

To verify the effectiveness of our proposed beamforming model, an experimental test is carried out. Fig. \ref{f-test}a illustrates the indoor test system and Fig. \ref{f-test}b is the on-site photos during the test. A high-frequency active antenna unit (AAU) is placed at the centerline of a corridor while a RIS is set up near the door of an empty room. The RIS, developed by ZTE, consists of 64$\times$64  4-bit and dual-polarized elements with the element size of 5 mm$\times$5 mm (Fig. \ref{f-test}b). A dedicated UE moves in the empty room along a circular trail for recording the signal. The AAU, RIS and UE are at the same height and the working frequency is 27.03708 GHz. The target reflection beam is set at $\theta_r=48^{\circ}$, $\varphi_r=180^{\circ}$. In order to perform numerical simulation for this test, the received power at the RIS element $P_{r}$ is measured as the reference signal receiving power (RSRP) at the surface of the RIS using the same UE, which is about -63 dBm in the test. The reflection amplitude of the RIS elements is 0.9, which is obtained from the full-wave EM simulation. It is difficult to measure the exact reflection pattern of a RIS element in practice and two typical patterns, $F(\theta,\varphi)=1$ and $F(\theta,\varphi)=\cos^{2}\theta$, are chosen for the simulation.

The measured data in Fig. \ref{f-test}c shows a narrow beam centered at $\theta=48.3^{\circ}$ and several ``side lobes'' also appear within the observation range. By moving the wave-absorbing board along the corridor, we find out that these ``side lobes'' are in fact multi-path signals reflected by the RIS. The minor fluctuation of the moving speed of the UE should account for the 0.3$^{\circ}$ offset of the target beam and the slight mismatch of the beamwidth between the measured data and the simulation results. The $\cos^{2}\theta$ pattern shows a higher gain (about 3 dB) at $\theta=48^{\circ}$ than the measured data while the isotropic one matches better, but it is not necessarily that the isotropic model fits the reflection pattern of the RIS element better since the gain deviation may result from the imperfection of the RIS or the inaccurate measurement of the RSRP. As to a beamforming model, the match of the beamwidth with the measured data is far more important than the power gain alignment. Hence, the comparison has verified that our reflection model is accurate enough for further numerical analysis.
\subsection{Phase-shift Error Model}
Assume that the elements of a RIS can be divided into $V$ groups with the $v$-th group having $N_v$ elements, $v=1,2,\cdots,V$, and $N=\sum_{v=1}^{V}N_v$, and that the phase-shift errors of the RIS elements within the same group can be modeled as i.i.d variables. By defining the phase-shift error of the $\ell$-th element in the $v$-th group $\Delta\phi_{v\ell}$ as the difference between the actual phase-shift $\phi_{v\ell}$ of the element and the optimal phase-shift $\phi_{v\ell}^{*}$ determined by \eqref{eq-phi-opt},
\begin{equation}
\Delta\phi_{v\ell}= \phi_{v\ell}-\phi_{v\ell}^{*},
\end{equation}
the received power in \eqref{eq-p-nf} can be rewritten as
\begin{equation}
P(\mathbf{r'})=S\left|\sum_{v=1}^{V}\sum_{\ell=1}^{N_{v}}\alpha_{v\ell}e^{j\Delta\phi_{v\ell}}\right|^2,\label{eq-p-group}
\end{equation}
where
\begin{equation}
\alpha_{v\ell}=\frac{\sqrt{P_{r,v\ell}F_{n}(\theta_{v\ell},\varphi_{v\ell})}}{|\mathbf{r}_{v\ell}-\mathbf{r}'|}|\Gamma_{v\ell}|e^{j\left(\phi_{v\ell}^{inc}+k|\mathbf{r}_{v\ell}-\mathbf{r}'|+\phi_{v\ell}^{*}\right)}
\end{equation}
is the \emph{amplitude attenuation factor} in free space. The received power at the target direction is of interest in the subsequent discussion and the parameter $\mathbf{r}'$ in $P(\mathbf{r}')$ will be omitted for simplicity, unless otherwise noted. 

In this paper, we focus on studying the effectiveness of an RIS between the transmitter and receiver, as also shown in our simulation setup in Fig. 3. Thus, we consider far-field propagation for RIS involved links. Within the far-field region, the following approximation can be made, $|\mathbf{r}_{v\ell}-\mathbf{r}'|\approx|\mathbf{r}'|+\frac{D}{2}\sin\theta_{r}$ \cite{balanis}, $\theta_{v\ell}\approx\theta_{r}$, $\varphi_{v\ell}\approx\varphi_{r}$, $\phi_{v\ell}^{\text{inc}}\approx\phi_{c}^{\text{inc}}$ and $P_{r,v\ell}\approx P_{r,c}$, where $\phi_{c}^{\text{inc}}$ and $P_{r,c}$ are the incident phase and  the received power at the central RIS element and $(\theta_r,\varphi_r)$ is the reflection angle with respect to the center of the RIS. Assuming that the reflection amplitude is a constant $|\Gamma|$ for all elements, the amplitude attenuation factor then becomes
\begin{equation}
\alpha_{v\ell}=\frac{\sqrt{P_{r,c}F_{n}(\theta_{r'},\varphi_{r'})}}{|\mathbf{r}'|+\frac{D}{2}\sin\theta_{r}}|\Gamma|e^{j\phi_{c}}=\alpha,\label{eq-a-2}
\end{equation}
where $\alpha$ is a complex constant for a given direction $(\theta_{r},\varphi_{r})$.
There are cases that the UE may appear in the near-field region when the RIS is extremely large \cite{cuimy}. However, the impact of the practical phase-shift errors on the beamforming gain may be alleviated due to the relatively low path loss in near-field region and hence the near-field cases are not discussed in this paper.

By applying the far-field approximation \eqref{eq-a-2}, the expectation of \eqref{eq-p-group} becomes
\begin{equation}
\begin{aligned}
\mathbb{E}[P]=&\mathbb{E}\Bigg[S\Big|\sum_{v=1}^{V}\sum_{\ell=1}^{N_{v}}\alpha e^{j\Delta\phi_{v\ell}}\Big|^2\Bigg]\\
=&S|\alpha|^2\mathbb{E}\Bigg[\Big|\sum_{v=1}^{V}\sum_{\ell=1}^{N_{v}}\cos(\Delta\phi_{v\ell})+j\sum_{v=1}^{V}\sum_{\ell=1}^{N_{v}}\sin(\Delta\phi_{v\ell})\Big|^2\Bigg]\\
=&S|\alpha|^2\Bigg[N+\sum_{v=1}^{V}\sum_{\ell=1}^{N_{v}}\sum_{m\neq\ell}^{N_{v}}\xi_{v}^{\ell m}+\sum_{v=1}^{V}\sum_{u\neq v}^{V}\sum_{\ell=1}^{N_{v}}\sum_{m=1}^{N_{u}}\zeta_{vu}^{\ell m}\Bigg],
\end{aligned}\label{eq-ep}
\end{equation}
where $\xi_{v}^{\ell m}=\mathbb{E}[\cos(\Delta\phi_{v\ell}-\Delta\phi_{vm})]$ and $\zeta_{vu}^{\ell m}=\mathbb{E}[\cos(\Delta\phi_{v\ell}-\Delta\phi_{um})]$. Since the phase-shift errors of the elements in the same group are i.i.d variables, i.e., $\mathbb{E}[\Delta\phi_{v\ell}]=\mathbb{E}[\Delta\phi_{vm}]$, the expectation $\mathbb{E}[\cos(\Delta\phi_{v\ell}-\Delta\phi_{vm})]$ has the same value for any $(\ell,m)$ pair as long as $\ell\neq m$, and so does the expectation $\mathbb{E}[\cos(\Delta\phi_{v\ell}-\Delta\phi_{um})]$. Hence, we can drop the superscript $\ell m$ in $\xi_{v}^{\ell m}$ and $\zeta_{vu}^{\ell m}$ and name $\xi_v$ the \emph{inner match factor} of group $v$ and $\zeta_{vu}$ the \emph{cross match factor} between the $v$-th and $u$-th groups. Equation \eqref{eq-ep} now can be simplified as
\begin{equation}
\mathbb{E}[P]=S|\alpha|^2\Bigg[N+\sum_{v=1}^{V}N_{v}(N_v-1)\xi_{v}+\sum_{v=1}^{V}\sum_{u\neq v}^{V}N_{v}N_{u}\zeta_{vu}\Bigg],\label{eq-ep-2}
\end{equation}
with
\begin{equation}
\left\{
\begin{aligned}
&\xi_{v}=\mathbb{E}[\cos\Delta\phi_{v}]^2+\mathbb{E}[\sin\Delta\phi_{v}]^2\\
&\zeta_{vu}=\mathbb{E}[\cos\Delta\phi_{v}]\mathbb{E}[\cos\Delta\phi_{u}]+\mathbb{E}[\sin\Delta\phi_{v}]\mathbb{E}[\sin\Delta\phi_{u}],
\end{aligned}\right.\label{eq-xi-zeta}
\end{equation}
where $\Delta\phi_{v}$ and $\Delta\phi_{u}$ represent the phase-shift error distribution of the $v$-th and $u$-th group, respectively. By applying \eqref{eq-ep-2}, phase-shift errors caused by various HWI can be investigated from the beamforming point of view.

\textbf{Proposition 1}. The beamforming gain loss due to the phase-shift errors of the RIS elements can be determined as
\begin{equation}
\delta=\frac{1}{N^2}\Bigg[N+\sum_{v=1}^{V}N_{v}(N_v-1)\xi_{v}+\sum_{v=1}^{V}\sum_{u\neq v}^{V}N_{v}N_{u}\zeta_{vu}\Bigg],\label{eq-delta}
\end{equation}
and it is independent of the incident angle of the EM wave if the phase-shift errors are randomly distributed.

\begin{IEEEproof}
If there is no phase-shift error on the RIS elements, then $\xi_{v}=\zeta_{vu}=1$ for $v,u\in\{1,2,\cdots,V\}$, which means that the phase delays of the signals reflected by each RIS element are perfectly matched at UE and the maximum power is received, which is

\begin{equation}
P_{\text{max}}=S|\alpha|^2\left[N+\sum_{v=1}^{V}N_{v}(N_v-1)+\sum_{v=1}^{V}\sum_{u\neq v}^{V}N_{v}N_{u}\right]=SN^2|\alpha|^2. \label{eq-p-max}
\end{equation}
According to \eqref{eq-ep-2} and \eqref{eq-p-max}, the ratio between $\mathbb{E}[P]$ and $P_{\text{max}}$ yields \eqref{eq-delta}. This ratio depicts the gain loss caused by phase-shift errors compared to the optimal beamforming gain. It is clear that $\delta$ is relevant to the number of elements and the distribution of the phase-shift errors in each group, but is independent of the incident angle of the EM wave since the incident angle only affects $\alpha$ but it is canceled out in \eqref{eq-delta}. 
\end{IEEEproof}
With this proposition, the path loss model between the BS and the RIS in \eqref{eq-p-inc} can be simply replaced by a given incident power in the numerical simulations when the gain loss due to phase-shift errors is of concern.
According to \eqref{eq-xi-zeta}, we have $\xi_v\leq 1$, $\zeta_{vu}\leq 1$, the dB value of $\delta$ is then non-positive and it represents directly the beamforming gain loss.
Hence, $\delta$ will be adopted as the key indicator for analyzing the impact of the practical phase-shift errors and its dB value is preferred in the following discussion.
\section{Impact of Phase-shift errors}\label{sec-3}
\subsection{Globally i.i.d Phase-shift Errors}\label{sec-iid}
The phase-shift errors of the RIS elements satisfying the same random distribution, i.e., globally i.i.d phase-shift errors,  is a special case of \eqref{eq-ep-2}. It has been stressed in some papers as listed in Section \ref{sec-intro} and we would like to further probe into this case.

\textbf{Proposition 2.} The power gain loss due to globally i.i.d phase-shift error is determined by the the inner match factor $\xi_{1}$ when $N\gg 1$.
\begin{IEEEproof}
For the globally i.i.d case, $V=1$, then \eqref{eq-ep-2} is reduced to a simpler form,
\begin{equation}
\mathbb{E}[P]=S|\alpha|^2[N+N(N-1)\xi_{1}].\label{eq-ep-giid}
\end{equation}
Clearly, the ratio between $\mathbb{E}[P]$ and $P_{\text{max}}$ is $\xi_{1}$ if $N\gg 1$, according to \eqref{eq-p-max} and \eqref{eq-ep-giid}, which completes the proof.
\end{IEEEproof}
There are two typical phase-shift error distributions, (1) uniformly distributed and (2) truncated normally distributed errors on RIS elements. The former is usually a consequence of the quantization of the phase-shift or the use of group control, i.e., a group of RIS elements is connected to a single input. The latter is mainly caused by fabrication errors, especially when all the elements of a RIS are fabricated in the same assemble line, following the same production procedure and quality control. A more general case is that the real distribution of the phase-shift errors is the mixture of two or more independent distributions.

\textbf{Remark 1.} If the phase-shift errors are uniformly distributed in $[\beta_1,\beta_2]$, i.e., $\Delta\phi\sim \mathcal{U}(\beta_{1},\beta_{2})$, we have
\begin{equation}
\xi_1(\Delta\phi)=\frac{2[1-\cos(\beta_{2}-\beta_{1})]}{(\beta_{2}-\beta_{1})^2}.\label{xi-u}
\end{equation}
Equation \eqref{xi-u} is obtained by the integral $\int_{\beta_1}^{\beta_2}\int_{\beta_1}^{\beta_2}\frac{\cos(x_1-x_2)}{\beta_2-\beta_1}dx_1dx_2$.
It is clear that $\xi_{1}(\Delta\phi)$ depends on the absolute error interval $(\beta_{2}-\beta_{1})$, rather than the exact error boundaries $\beta_{1}$ and $\beta_{2}$. That is,  for a uniform distribution $\Delta\phi\sim\mathcal{U}(\beta_{1}+\eta,\beta_{2}+\eta)$, $\xi_{1}(\Delta\phi)$ has the same value for any  $\eta\in\mathbb{R}$. The offset $\eta$ makes no difference to the beamforming gain, which can be easily proved as following,
\begin{equation}
\Big|\sum_{\ell=1}^{N}\alpha_{\ell}e^{j(\Delta\phi_{\ell}+\eta)}\Big|=\Big|e^{j\eta}\sum_{\ell=1}^{N}\alpha_{\ell}e^{j\Delta\phi_{\ell}}\Big|
=\Big|\sum_{\ell=1}^{N}\alpha_{\ell}e^{j\Delta\phi_{\ell}}\Big|.\label{eq-eta-offset}
\end{equation}
Consequently, $\Delta\phi\sim\mathcal{U}(\beta_{1},\beta_{2})$ is equivalent to a zero-mean distribution $\Delta\phi\sim\mathcal{U}(-\frac{\beta_2-\beta_1}{2},\frac{\beta_2-\beta_1}{2})$ with respect to the beamforming gain.

\textbf{Remark 2. }If the phase-shift errors satisfy a truncated normal distribution with mean value $\mu$, standard deviation $\sigma$ and truncation interval $[\mu-\psi,\mu+\psi]$, $\Delta\phi\sim\mathcal{N}_{\text{t}}(\mu,\sigma,\mu-\psi,\mu+\psi)$, we have
\begin{equation}
\xi_{1}(\Delta\phi)=e^{-\sigma^2}\Bigg[\frac{\text{erf}\left(\frac{\psi-j\sigma^{2}}{\sqrt{2}\sigma}\right)+\text{erf}\left(\frac{\psi+j\sigma^{2}}{\sqrt{2}\sigma}\right)}{2 \text{erf}\left(\frac{\psi}{\sqrt{2}\sigma}\right)}\Bigg]^{2},
\label{gamma-n}
\end{equation}
where $\text{erf}(z)=2/\sqrt{\pi}\int_{0}^{z}e^{-t^2}dt$ is the error function.
It can be found that $\xi_{1}(\Delta\phi)$ relies on the error deviation $\sigma$ and the absolute truncation interval [$-\psi$,$\psi$] centered at $\mu$ but is independent of $\mu$. The mean value $\mu$ can as well be treated as a common offset, the same as $\eta$ in \eqref{eq-eta-offset}, and thus $\Delta\phi\sim \mathcal{N}_{\text{t}}(\mu,\sigma,\mu-\psi,\mu+\psi)$ is equivalent to $\Delta\phi\sim \mathcal{N}_{\text{t}}(0,\sigma,-\psi,\psi)$ in this case.
\begin{figure}[h]
\centering{\includegraphics[width=0.96 \textwidth]{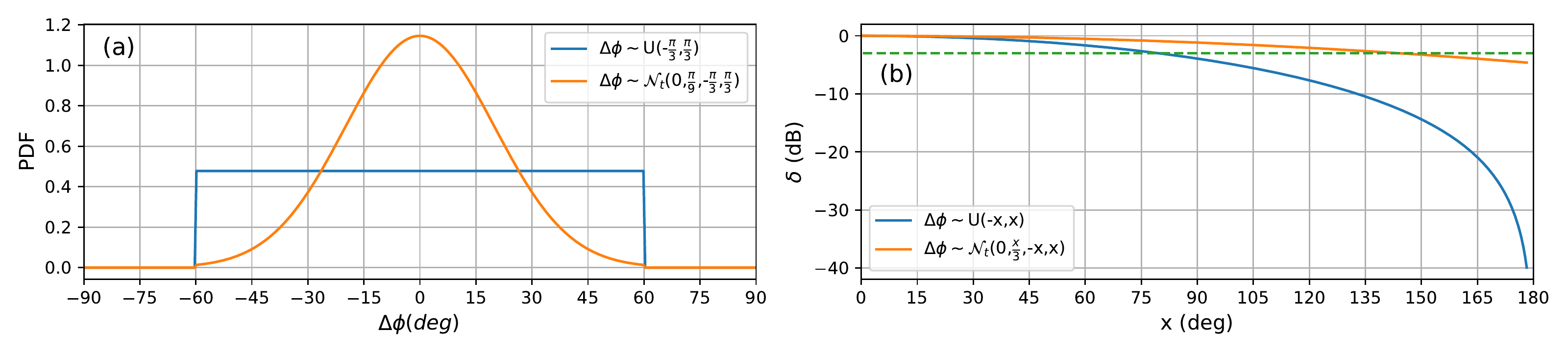}}
\caption{(a) PDF examples of uniform distribution and truncated normal distribution. (b) Power gain loss expectation under uniform and truncated normal distributions. The green dashed line marks the -3 dB level.}
\label{f-pdf}
\end{figure}

Examples of the probability density function (PDF) of uniform distribution and truncated normal distribution are shown in Fig. \ref{f-pdf}a. The truncation interval of the truncated normal distribution is chosen to be the 3-sigma region $[-3\sigma,3\sigma]$. Fig. \ref{f-pdf}b shows the beamforming gain loss $\delta$ of uniform distribution and truncated normal distribution versus the error range. Clearly, $\delta$ drops quickly when the error range is larger than $[-\frac{\pi}{4},\frac{\pi}{4}]$ for the uniform distribution while it declines slower for the truncated normal distribution. A 3 dB loss is witnessed when the phase-shift errors satisfy $\Delta\phi\sim\mathcal{U}(-79^{\circ},79^{\circ})$ or $\Delta\phi\sim\mathcal{N}_{\text{t}}(0,48^{\circ},-144^{\circ},144^{\circ})$. The good news is that when the phase-shift error is confined in a small range, feeble gain loss is expected. For example, when the errors are limited in $[-\frac{\pi}{6}$,$\frac{\pi}{6}]$, the losses are 0.40 dB and 0.13 dB for a uniform distribution and a truncated normal distribution, respectively. When $\Delta\phi\sim\mathcal{U}(-\frac{\pi}{2},\frac{\pi}{2})$, it is equivalent to a 1-bit phase-shift quantization and $\xi_1=\frac{4}{\pi^2}\approx 3.9 $ dB, the same as in \cite{wuqq2}.

\begin{figure}[h]
\centering
\subfloat{
\includegraphics[width=0.8 \textwidth]{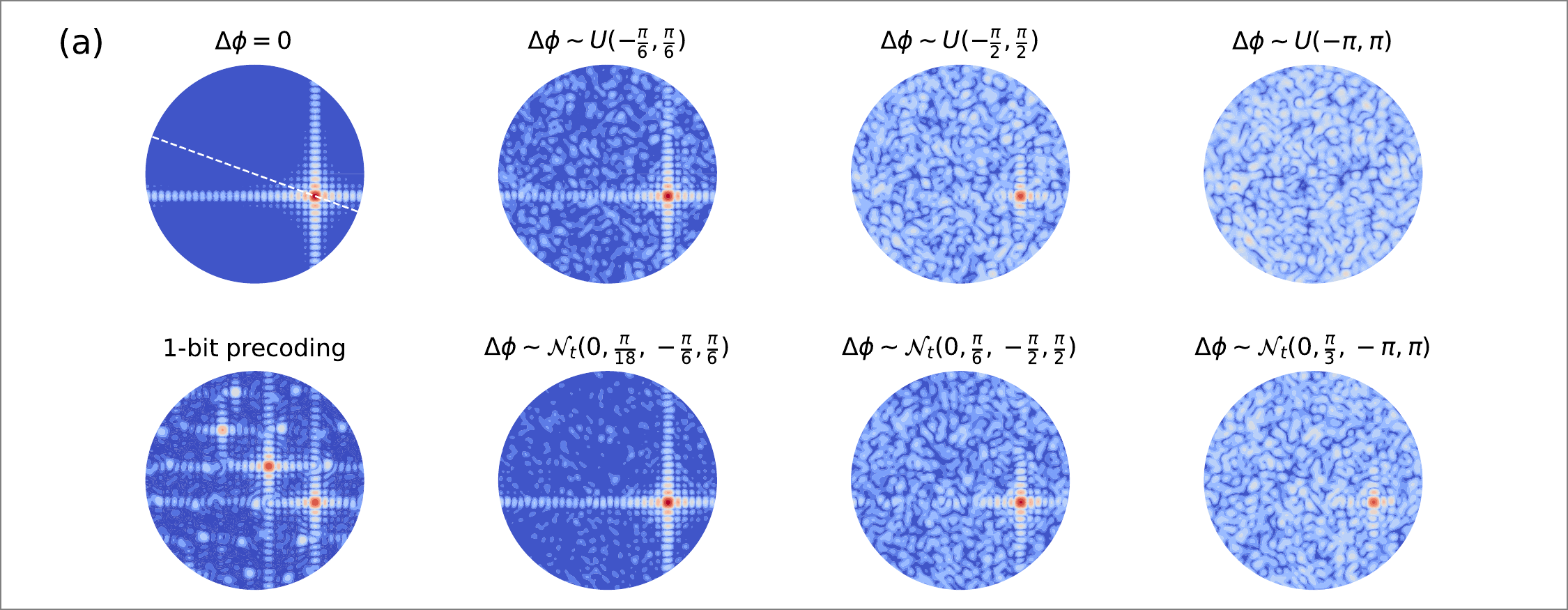}}\vspace{3mm}\;
\subfloat{
\includegraphics[width=0.98 \textwidth]{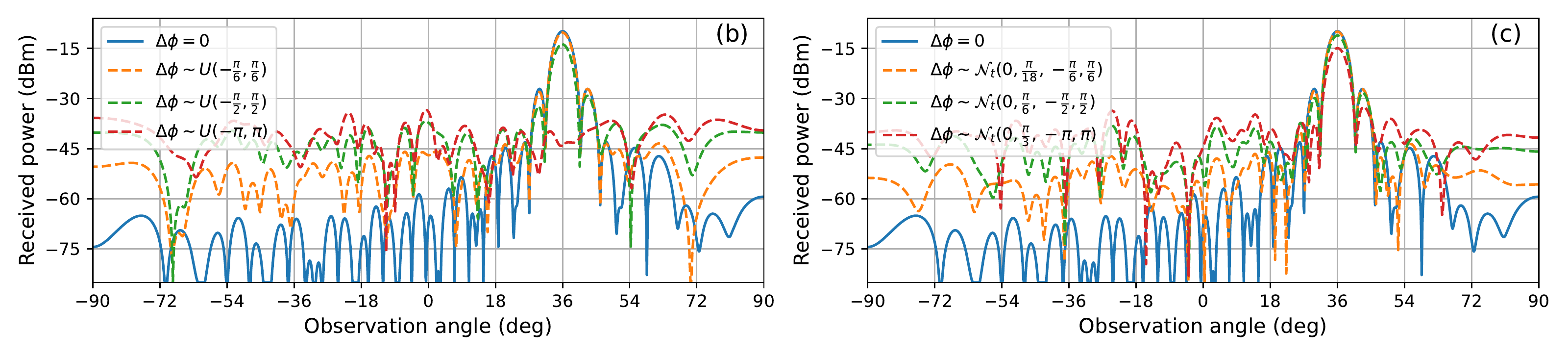}}
\caption{The 3D patterns on the hemispherical surface (a) and the 2D beam patterns under uniformly distributed errors (b) and truncated normally distributed errors (c).   Note that the 3D patterns on the hemispherical surface have been projected onto the x-o-y plane and the same color scale is applied. }
\label{f-pat2d-nu}
\end{figure}
Fig. \ref{f-pat2d-nu} shows the numerical results of different error distributions with simulation parameters listed in Table \ref{t-sim1-param}. The power level of the side lobes goes up notably under phase-shift errors as indicated in Fig. \ref{f-pat2d-nu}a. The yellow dashed line in the upper left contour in Fig. \ref{f-pat2d-nu}a marks the orthodrome that the 2D pattern locates. As expected, no evident power gain reduction is observed at the target direction when the randomly distributed phase-shift errors are confined in the range $[-\frac{\pi}{6}$,$\frac{\pi}{6}]$ for either uniform distribution or truncated normal distribution (the dashed orange line in Fig. \ref{f-pat2d-nu}b and Fig. \ref{f-pat2d-nu}c) while no main lobe appears at the target direction when the phase-shift errors uniformly distributed in $[-\pi$,$\pi]$ (the dashed red line in Fig. \ref{f-pat2d-nu}b). Although it has been mentioned that the 1-bit precoding is equivalent to the phase-shift errors distributed in $[-\frac{\pi}{2},\frac{\pi}{2}]$ uniformly with respect to the power gain at the UE, the 3D beam patterns are totally different for these two cases. The former gives rise to a grating lobe at a different direction while the latter spreads the power onto the whole hemispherical surface, as shown in Fig. \ref{f-pat2d-nu}a. The increase of power level of the side lobes due to phase-shift errors may raise up the interference level to other users.
\begin{table}
\caption{Parameters for the simulation in Fig. \ref{f-pat2d-nu}}
\label{t-sim1-param}
\centering
\begin{tabular}{|l|l|}
\hline
Parameter&Value\\
\hline
Carrier frequency&27.03708 GHz\\
Array size & $N_x=N_y=32$ \\
Element size&$d_x=d_y=0.5\lambda$\\
Location of BS&$d_t=100 \text{m},\theta_t=20^{\circ},\varphi_{t}=174^{\circ}$\\
Location of UE&$d_r=10 \text{m},\theta_r=36^{\circ},\varphi_{r}=340^{\circ}$\\
Incident power at RIS elements&0 dBm\\
Reflection amplitude&$|\Gamma|=1.0$\\
Element reflection pattern&$F(\theta,\varphi)=1$ (isotropic)\\
Effective receiving aperture&$S=d_x*d_y$\\
\hline
\end{tabular}
\end{table}

When a RIS encounters more than one adverse condition during the fabrication or deployment, the total phase-shift error of an element is the superposition of the errors aroused by each of these adverse conditions. For example, if uniformly distributed phase-shift errors in the range $[\beta_{1},\beta_{2}]$ are brought in in the fabrication stage of a 1-bit RIS, the phase-shift error consists of two distributions, $\Delta\phi^{1}\sim \mathcal{U}(-\frac{\pi}{2},\frac{\pi}{2})$ and  $\Delta\phi^{2}\sim \mathcal{U}(\beta_{1},\beta_{2})$. Similarly, if the fabrication error satisfies a truncated normal distribution, the total phase-shift error on each element then consists of both uniformly distributed and truncated normally distributed components. 

\textbf{Proposition 3. }For a group of RIS elements having hybrid phase-shift errors, if
\begin{itemize}
\item the phase-shift error is composed of $n$ error components, $\Delta\phi=\sum_{i=1}^{n}\Delta\phi^i$,
\item $\Delta\phi^i$ satisfies a random distribution $\mathcal{D}_i$ defined in $(a_i, b_i)$ and the PDF of $\mathcal{D}_i$ is symmetric with respect to $\Delta\phi^i=(a_i+b_i)/2$ for $i=1,2,\cdots,n$,
\end{itemize}
 then the following relation holds
\begin{equation}
\xi_{1}(\Delta\phi)=\prod_{i=1}^{n}\xi_{1}(\Delta\phi^i).
\end{equation}
\begin{IEEEproof}
Detailed proof is provided in Appendix.
\end{IEEEproof}
\subsection{Grouped Random Phase-shift Errors}
As is discussed previously, randomly distributed phase-shift errors may be introduced during fabrication and different error types may emerge on different group of elements. Such situation may occur when a metasurface consists of different types of elements or when the state of the RIS elements can be reconfigured. An example is the dog-bone shaped metallic unit cells designed in \cite{lavigne} for forming refraction metasurfaces. Since the shape of the patch differs from each other, different phase-shift errors may emerge due to fabrication error. Another example is the 2-bit element proposed in \cite{daill}, where the phase-shift state of the element is determined by the combination of the ON/OFF state of different PIN diodes and thus the phase-shift errors may differ in different element states.

The calculation of cross match factor $\zeta_{vu}$ becomes complicated when phase-shift error distributions differ among element groups, but we can still obtain the analytical expressions by commercial software, such as Mathematica,  as given in \eqref{eq-zeta-numu}.
\begin{figure*}[!t]
\setcounter{eqncnttmp}{\value{equation}}
\setcounter{equation}{23}
\begin{equation}
\zeta_{vu}=\left\{
\begin{aligned}
&\frac{\cos(\beta_{1}-\gamma_{1})+\cos(\beta_{2}-\gamma_{2})-\cos(\beta_{1}-\gamma_{2})-\cos(\beta_{2}-\gamma_{1})}{(\beta_{2}-\beta_{1})(\gamma_{2}-\gamma_{1})},\\&\qquad\Delta\phi_{v}\sim\mathcal{U}(\beta_{1},\beta_{2}),\Delta\phi_{u}\sim\mathcal{U}(\gamma_{1},\gamma_{2});\\
&\frac{\sin(\beta_1-\mu)-\sin(\beta_2-\mu)}{(\beta_1-\beta_2)e^{-\frac{1}{2}\sigma^2}}\frac{\text{erf}\left(\frac{\psi-j\sigma^2}{\sqrt{2}\sigma}\right)+\text{erf}\left(\frac{\psi+j\sigma^2}{\sqrt{2}\sigma}\right)}{2\text{erf}\left(\frac{\psi}{\sqrt{2}\sigma}\right)},\\&\qquad\Delta\phi_{v}\sim\mathcal{U}(\beta_{1},\beta_{2}),\Delta\phi_{u}\sim\mathcal{N}_{\text{t}}(\mu,\sigma,\mu-\psi,\mu+\psi);\\
&\frac{\cos(\mu_1-\mu_2)}{e^{\frac{1}{2}(\sigma_{1}^{2}+\sigma_{2}^{2})}}\frac{\left[\text{erf}\left(\frac{\psi_1-j\sigma_{1}^{2}}{\sqrt{2} \sigma_{1}}\right)+\text{erf}\left(\frac{\psi_1+j\sigma_{1}^{2}}{\sqrt{2} \sigma_{1}}\right)\right] \left[\text{erf}\left(\frac{\psi_2-j\sigma_{2}^{2}}{\sqrt{2} \sigma_{2}}\right)+\text{erf}\left(\frac{\psi_2+\sigma_{2}^2}{\sqrt{2} \sigma_{2}}\right)\right]}{4\text{erf}\left(\frac{a}{\sqrt{2}\sigma_{1}}\right)\text{erf}\left(\frac{\psi_2}{\sqrt{2}\sigma_{2}}\right)},\\
&\qquad\Delta\phi_{v}\sim\mathcal{N}_{\text{t}}(\mu_{1},\sigma_{1},\mu_1-\psi_1,\mu_1+\psi_1),\Delta\phi_{u}\sim\mathcal{N}_{\text{t}}(\mu_{2},\sigma_{2},\mu_2-\psi_2,\mu_2+\psi_2).
\end{aligned}\label{eq-zeta-numu}
\right.
\end{equation}
\hrulefill
\vspace*{4pt}
\setcounter{equation}{24}
\end{figure*}
Carrier frequency mismatch and PIN diode failure are common issues in RIS application and grouped i.i.d phase-shift errors occur in both issues. The beamforming gain loss of these two issues will be discussed by means of  the phase-shift error model \eqref{eq-ep} in this section.
\begin{figure}[h]
\centering{\includegraphics[width=0.45 \textwidth]{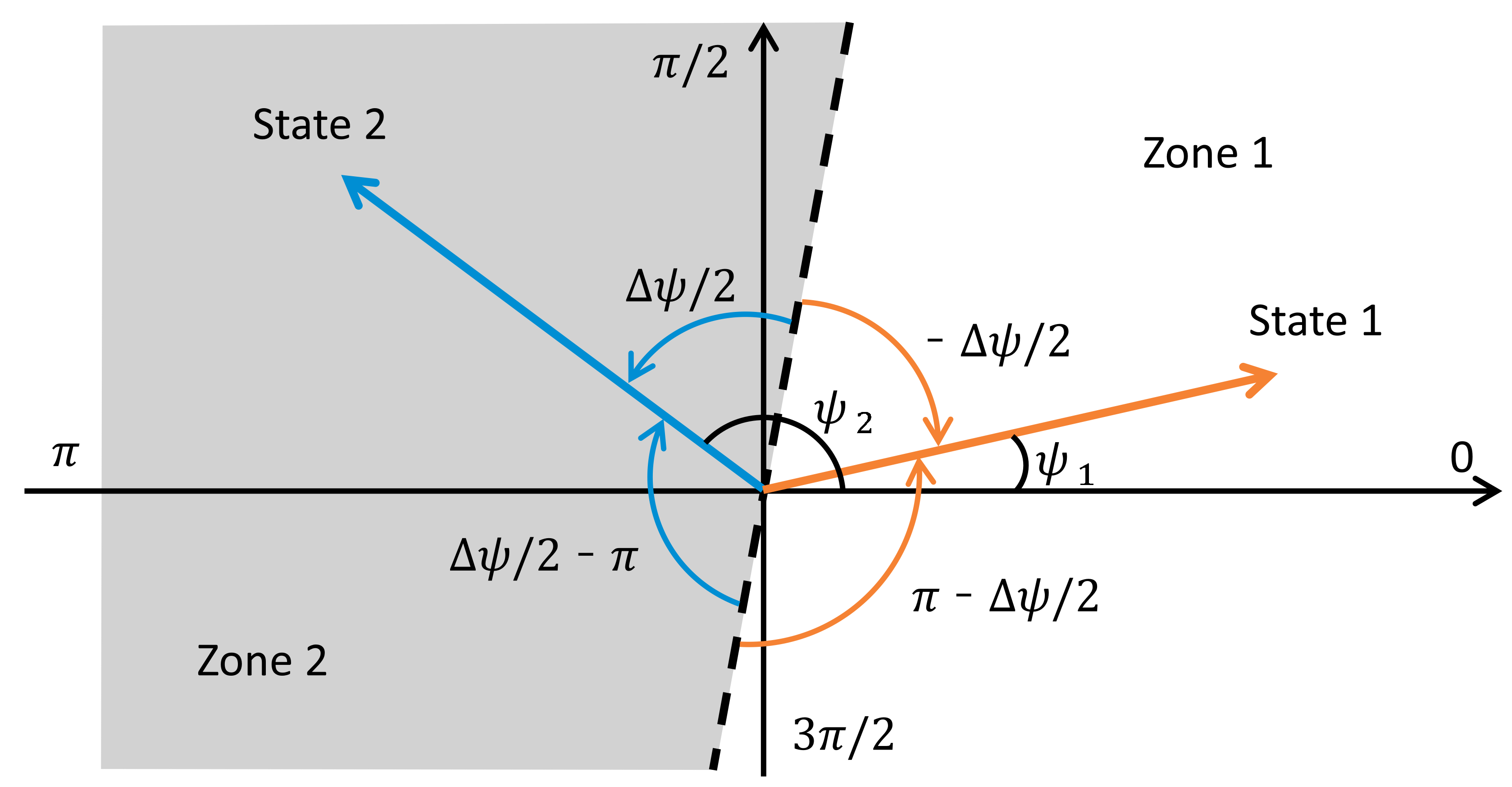}}
\caption{Diagram of the phase-shift error distributions of a 1-bit RIS element.}
\label{f-phi-1bit}
\end{figure}
\subsubsection{Carrier Frequency Mismatch}
Usually, the RIS element has finite reconfigurable states with each state corresponding to a specific phase-shift and the phase-shift of a certain state varies with the frequency of the impinging EM waves. For example, the 1-bit RIS element proposed in \cite{cuitj} has phase-shift difference between the two states ranging from 135$^{\circ}$ to 200$^{\circ}$ in the frequency band 8.1 GHZ to 12.7 GHz. Assuming that a 1-bit RIS element working at a certain frequency has phase-shifts $\psi_1$ at State 1 and $\psi_2$ at State 2 (Fig. \ref{f-phi-1bit}), then the state of an element in the beamforming are chosen according to the optimal phase-shift determined by \eqref{eq-phi-opt}. State 1 is chosen when the optimal phase-shift locates in Zone 1 and State 2 is chosen otherwise. Accordingly, the phase-shift errors at the two states satisfy $\Delta\phi_{1}\sim\mathcal{U}(\frac{\Delta\psi}{2}-\pi,\frac{\Delta\psi}{2})$ and $\Delta\phi_{2}\sim\mathcal{U}(-\frac{\Delta\psi}{2},\pi-\frac{\Delta\psi}{2})$ with $\Delta\psi=\psi_2-\psi_1$, which yields
\begin{equation}
\delta=\frac{1}{N^2}\left(N_1^2\xi_1+N_2^2\xi_2+2N_1N_2\zeta_{12}\right).\label{eq-dg-1bit-fm}
\end{equation}

\textbf{Remark 3. }Statistically, $N_1=N_2=\frac{N}{2}$ provided that $N$ is sufficient large. By applying \eqref{xi-u} and \eqref{eq-zeta-numu}, we have $\xi_{1}=\xi_{2}=\frac{4}{\pi^2}$ and $\zeta_{12}=-\frac{4}{\pi^2}\cos(\Delta\psi)$, then \eqref{eq-dg-1bit-fm} becomes
\begin{equation}
\delta=\frac{2}{\pi^2}\left[1-\cos(\Delta\psi)\right],\label{eq-dg-1bit-fm2}
\end{equation}
which depicts the beamforming gain loss for a 1-bit RIS with elements having phase-shift difference $\Delta\psi$ between the 2 states. A maximum value $\frac{4}{\pi^2}$ is achieved in \eqref{eq-dg-1bit-fm2} when $\Delta\psi=\pi$, which indicates that the optimal phase-shift difference between the two states of a 1-bit element is $\pi$. In other words, the optimal working frequency of a 1-bit RIS is the frequency with which the phase-shift difference is $\pi$ between the two states. The reason is straightforward, as the $\pi$ difference between the two phase-shift partitions the $2\pi$ period into two equal portions and minimizes the maximum phase-shift error to $\pm \frac{\pi}{2}$ for both states. Similarly, the optimal phase-shift set for a $q$-bit RIS element should has the phase-shift difference $\frac{2\pi}{2^q}$ between the adjacent states. If the carrier frequency is not at the optimal frequency band, there is beamforming gain loss.
 \begin{figure}[h]
\centering{\includegraphics[width=0.98 \textwidth]{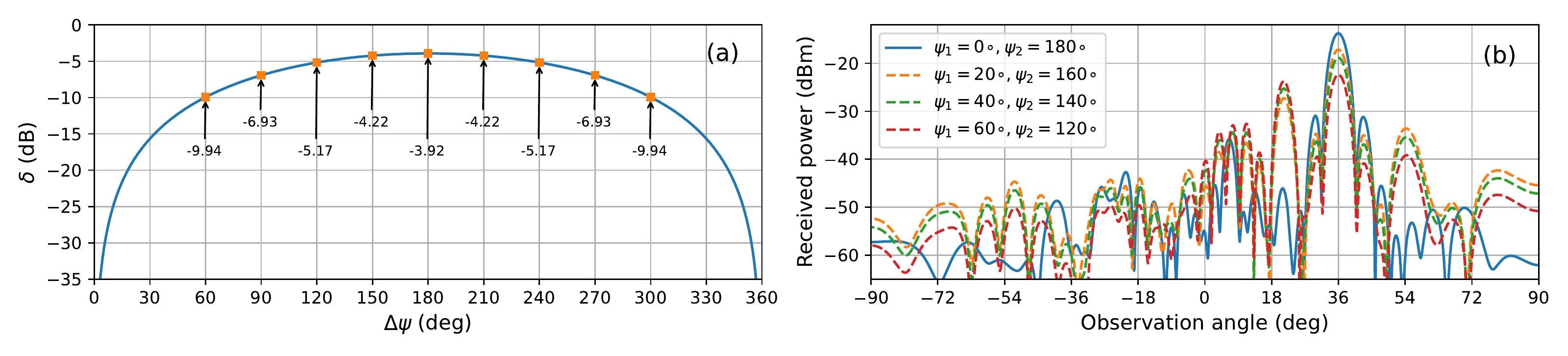}}
\caption{(a) Theoretical beamforming gain loss versus phase difference for 1-bit RIS. (b) Simulation result of 2D beam patterns with different 1-bit phase-shift set. The legend labels denote the 2 discrete phase-shifts in degree.}
\label{f-dg-1bit}
\end{figure}

The beamforming gain loss of a 1-bit RIS with different $\Delta\psi$ are plotted in Fig. \ref{f-dg-1bit}a. The gain only drops about 1.3 dB when the phase-shift difference $\Delta\psi$ deviates 60$^{\circ}$ from the optimal one (180$^{\circ}$), i.e., $\Delta\psi\in[120^{\circ},240^{\circ}]$, which is a relatively large range and may ease the design of the RIS element. However, the total gain loss rises sharply if $\Delta\psi$ keep deviating from 180$^{\circ}$ and a 10 dB drop is observed when $\Delta\psi$ shrinks to 60$^{\circ}$ or expands to 300$^{\circ}$. Fig. \ref{f-dg-1bit}b shows the numerical result of a 1-bit RIS with various $\Delta\psi$ under the parameters listed in Table \ref{t-sim1-param}, which also verifies that the beamforming gain drops gradually with the shrink of the phase-shift difference between the two states of the 1-bit RIS element. The simulated gain loss at the target direction is lower than the theoretical value 10 dB when $\Delta\psi=60^{\circ}$ and the reason may be that the 32$\times$32 element array is not large enough.

We need to further emphasize that the beamforming gain loss in Fig. \ref{f-dg-1bit}a is obtained under the assumption that the phase-shifts of the two states of the RIS element are known and the phase-shift matrix is designed based on these phase-shifts. However, if the actual phase-shift difference is unknown and the phase-shift matrix is designed based on the optimal one ($\Delta\psi=\pi$), larger power gain loss will be expected as the error range is unintentionally enlarged. On the other hand, if the RIS is utilized for wide-band communications, i.e. a single frequency precoding matrix for all sub-carriers, the beamforming gain loss will be further increased(e.g. \cite{wuk}) and this gain loss in fact involves in another type of phase-shift errors which will be discussed in Section \ref{sec-fixed-error}.
\subsubsection{PIN Diode Failure}
For the RIS with discrete phase-shift controlled by PIN diodes, the failure of PIN diode may occur gradually after a long period of deployment and result in undesired change to the element state, which is a serious issue to the RIS-assisted communication systems. There are usually  two types of PIN diode failure, open circuit and short circuit, which lead to permanent OFF and ON states and give rise to the open-circuit and short-circuit phase-shift errors, respectively. Take the 2-bit RIS element proposed in \cite{daill} for example, configuration 2 will switch to configuration 1 if PIN 1 is short-circuit and PIN 2 is open-circuit (Table 1 in \cite{daill}).

It should be pointed out that an element with a PIN diode having open-circuit failure is still functional if the element state required for the beamforming is the open-circuit state, and so does the element with PIN diode having short-circuit failure. Statistically, half of the elements with PIN diode failure can still work properly for beamforming. For a 1-bit RIS with $N$ elements, let $p_1$ and $p_2$ respectively be the ratio of the elements switched to State 1 and 2 permanently due to PIN diode failures, where $0\leq p_1\leq 1$, $0\leq p_2\leq 1$, and $p_1+p_2\leq 1$, then the RIS elements can be divided into four groups as in Table \ref{t-pin-fail}. The number of elements in each group can be estimated statistically and the phase-shift error distribution of each group can be determined according to Fig. \ref{f-phi-1bit}.
\begin{table*}
\caption{Element groups in a RIS with PIN diode failures}
\label{t-pin-fail}
\centering
\resizebox{\textwidth}{!}{
\begin{tabular}{|c|l|l|l|}
\hline
Group No.&Element type&Number of elements&Error distribution\\
\hline
1&\tabincell{l}{Elements switched to State 1 as expected for beamforming\\ and elements at State 1 permanently due to PIN failure}&$N_1=\frac{1}{2}(1-p_2)N$&$\Delta\phi_{1}\sim\mathcal{U}(\frac{-\Delta\psi}{2},\pi-\frac{\Delta\psi}{2})$\\
2&\tabincell{l}{Elements switched to State 2 as expected for beamforming\\ and elements at State 2 permanently due to PIN failure}& $N_2=\frac{1}{2}(1-p_1)N$&$\Delta\phi_{2}\sim\mathcal{U}(\frac{\Delta\psi}{2}-\pi,\frac{\Delta\psi}{2})$\\
3&\tabincell{l}{Elements at State 1 permanently due to PIN failure but \\State 2 is expected for beamforming}&$N_3=\frac{1}{2}p_1N$&$\Delta\phi_{3}\sim\mathcal{U}(\frac{\Delta\psi}{2},\frac{\Delta\psi}{2}+\pi)$\\
4&\tabincell{l}{Elements at State 2 permanently due to PIN failure but \\State 1 is expected for beamforming}&$N_4=\frac{1}{2}p_2N$&$\Delta\phi_{4}\sim\mathcal{U}(\frac{-\Delta\psi}{2}-\pi,-\frac{\Delta\psi}{2})$\\
\hline
\end{tabular}}
\end{table*}

\begin{figure}[h]
\centering{\includegraphics[width=0.48 \textwidth,trim=30 0 0 0,clip]{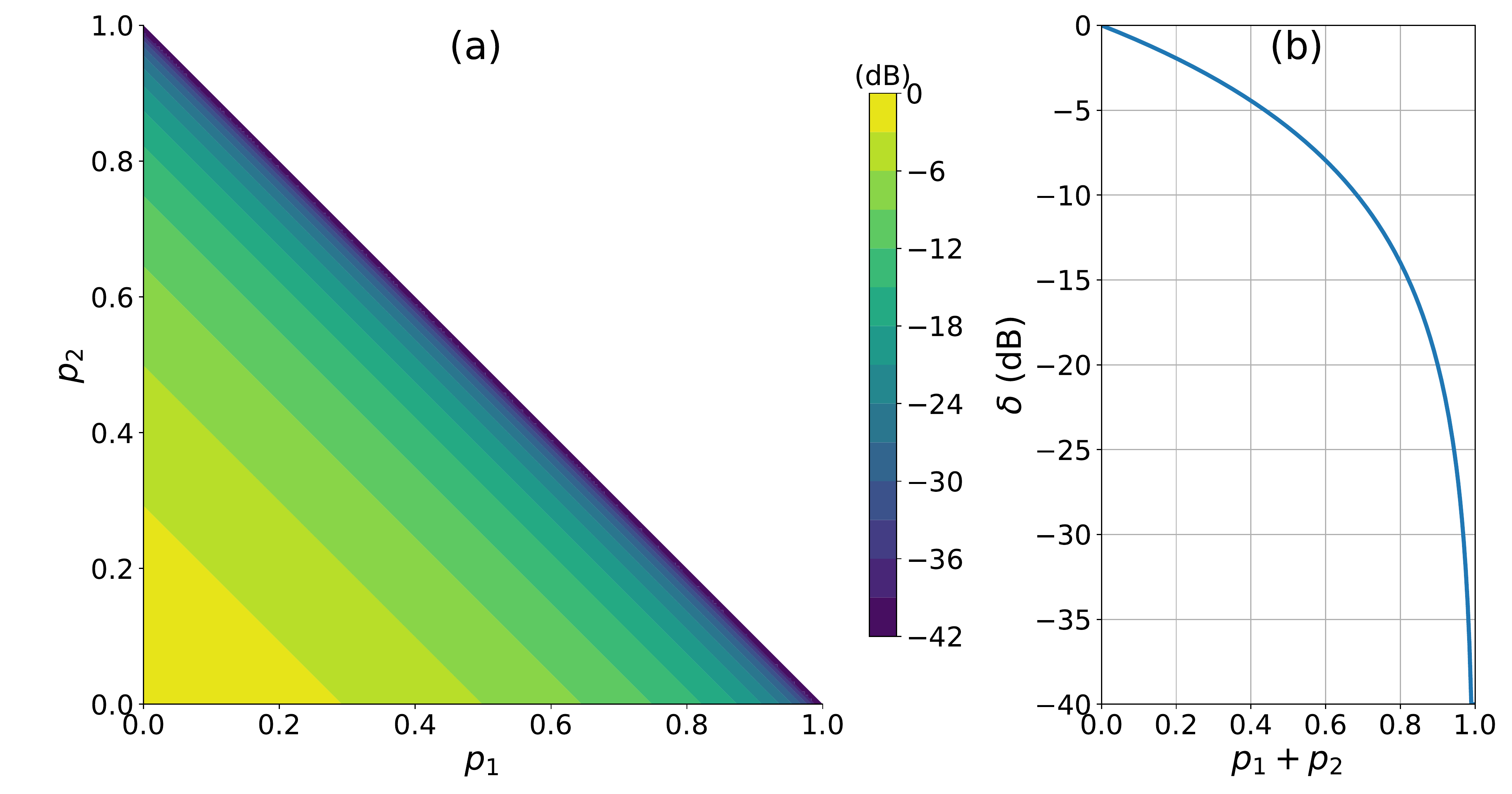}}
\caption{(a) Power gain loss contour with PIN diode failures. (b) Power gain loss profiles.}
\label{f-pin-fail}
\end{figure}

\textbf{Remark 4. }For the 4 groups of elements, we obtain
\begin{equation}
\delta=\frac{1}{N^2}\left[\sum_{v=1}^{4}N_{v}^2\xi_{v}+2\sum_{v=1}^{4}\sum_{u=v+1}^{4}N_{v}N_{u}\zeta_{vu}\right].\label{eq-dg-pin}
\end{equation}
The inner match factor and cross match factor can be determined by \eqref{xi-u} and \eqref{eq-zeta-numu}, which are $\xi_{1}=\xi_{2}=\xi_{3}=\xi_{4}=-\zeta_{14}=-\zeta_{23}=\frac{4}{\pi^2}$, $\zeta_{13}=\zeta_{24}=-\zeta_{12}=-\zeta_{34}=\frac{4\cos\Delta\psi}{\pi^2}$.
Then, the beamforming gain loss of 1-bit RIS with PIN diode failure becomes
\begin{equation}
\delta=\frac{1}{\pi^2}\left(t_1^2+t_2^2-2t_1t_2\cos\Delta\psi\right),\label{eq-dg-pin-2}
\end{equation}
where $t_1=1-2p_1$ and $t_2=1-2p_2$. The ratio $\delta$ reduces to \eqref{eq-dg-1bit-fm2} if $p_1=p_2=0$. Since the beamforming gain loss of optimal 1-bit quantization is $\frac{4}{\pi^2}$, let $\Delta\psi=\pi$, we obtain the pure beamforming gain loss caused by PIN diode failures, which is $\delta=(1-p_1-p_2)^2$. As is shown in Fig. \ref{f-pin-fail}, the gain loss depends on the total ratio $p_1+p_2$ and the loss is less than 3 dB if $p_1+p_2<0.29$. 
\begin{figure}[h]
\centering
\subfloat[]{
\includegraphics[width=0.25 \textwidth]{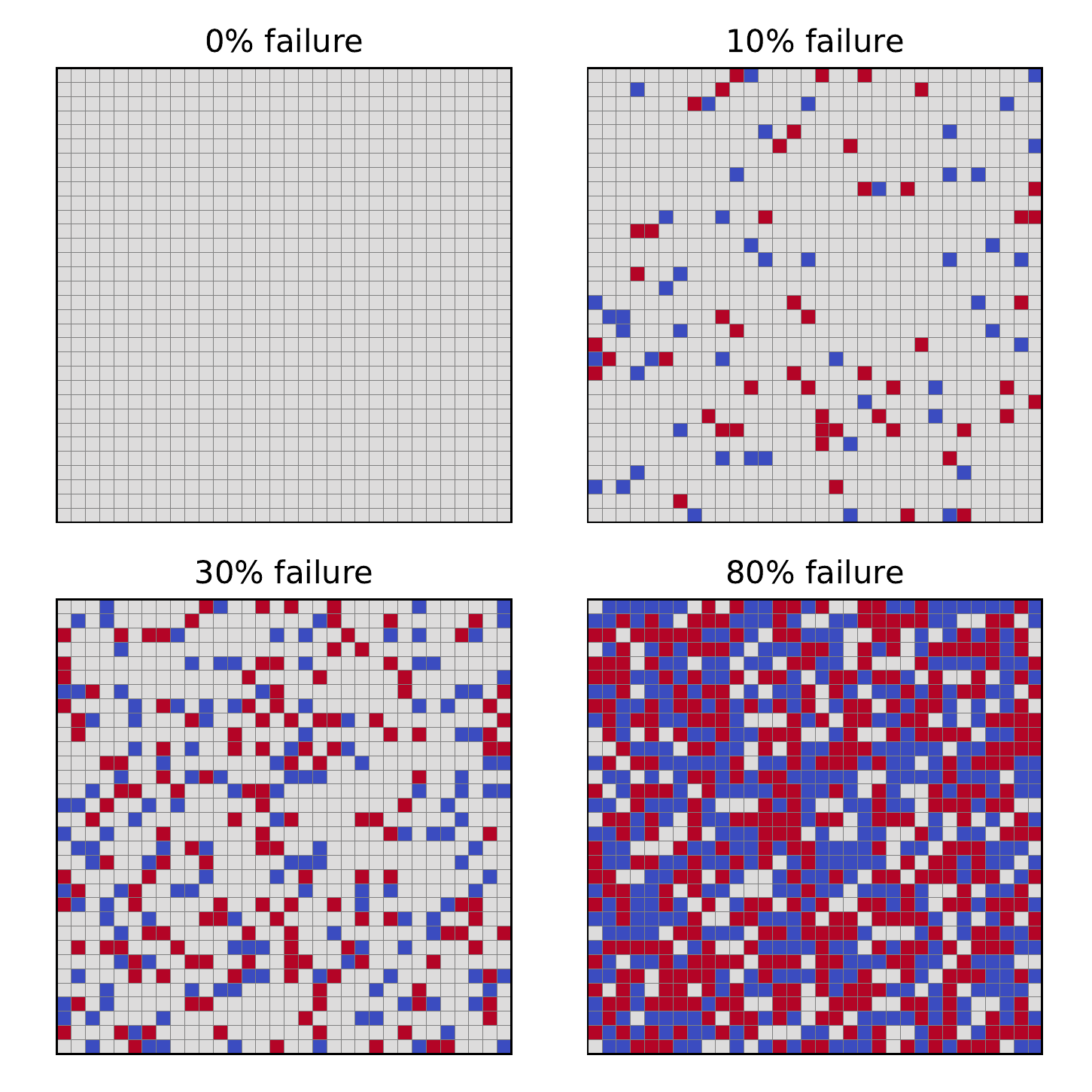}}\quad
\subfloat[]{
\includegraphics[width=0.46 \textwidth]{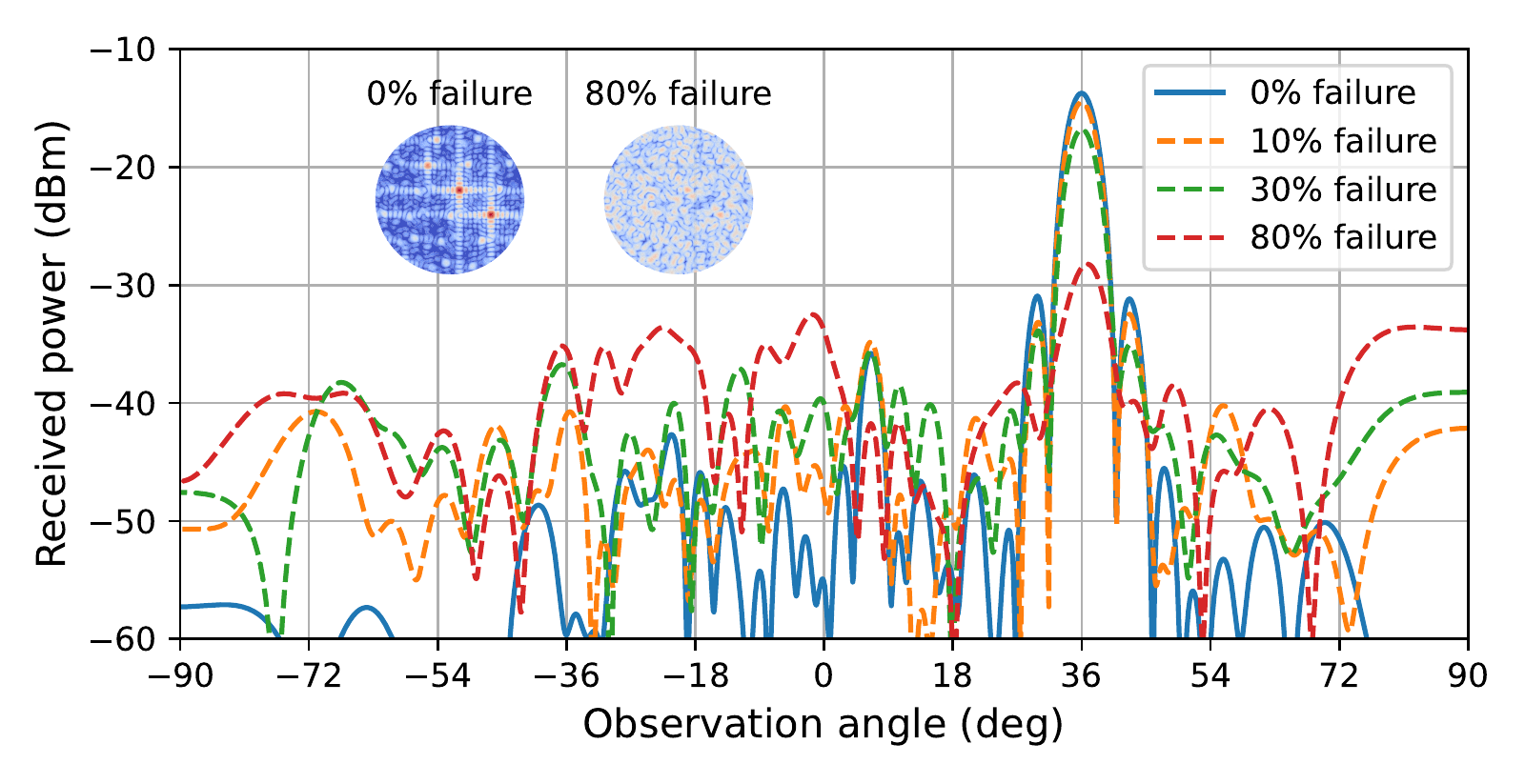}}
\caption{(a) Illustrations of PIN diode status where the dark red and blue patches stand for the open-circuit and short-circuit failures while the gray patches are the functional elements, and (b) simulated the beam pattern with PIN diode failures. An animated version of this figure is available in the supplimentary material.}
\label{f-pin-fail-sim}
\end{figure}

The simulation result in Fig. \ref{f-pin-fail-sim} reveals the change of the main lobe under different PIN diode failure proportions. The peak gain of the main lobe with 30\% PIN diode failure is about 2.9 dB lower than that without failure, which is close to the theoretical value. When 80\% of the PIN diodes are out of work, the reflection energy almost diffuses and the RIS is nothing but a rough surface.

If real-time beam tracking is required for a RIS, the PIN diodes embedded is going to be switched heavily between ON and OFF states and the lifespan of a RIS can be evaluated with the proposed phase-shift error model. The 1-bit example is presented in this subsection but RIS with higher quantization level can be analyzed once the phase-shift errors with PIN diode failures are determined.

\subsection{Grouped Fixed Phase-shift Errors}\label{sec-fixed-error}
The impact of randomly distributed phase-shift errors are discussed in previous sections but there are cases that a RIS is accompany by fixed phase-shift errors. The impact of two types of fixed phase-shift errors is to be analyzed in this section. The first type of fixed phase-shift error is introduced by the unwanted deformation of RIS panel and the second type usually occurs in wideband system.
\subsubsection{RIS Panel Deformation}
\begin{figure}[h]
\centering
\subfloat[]{
\includegraphics[width=0.26 \textwidth,trim=60 120 80 140,clip]{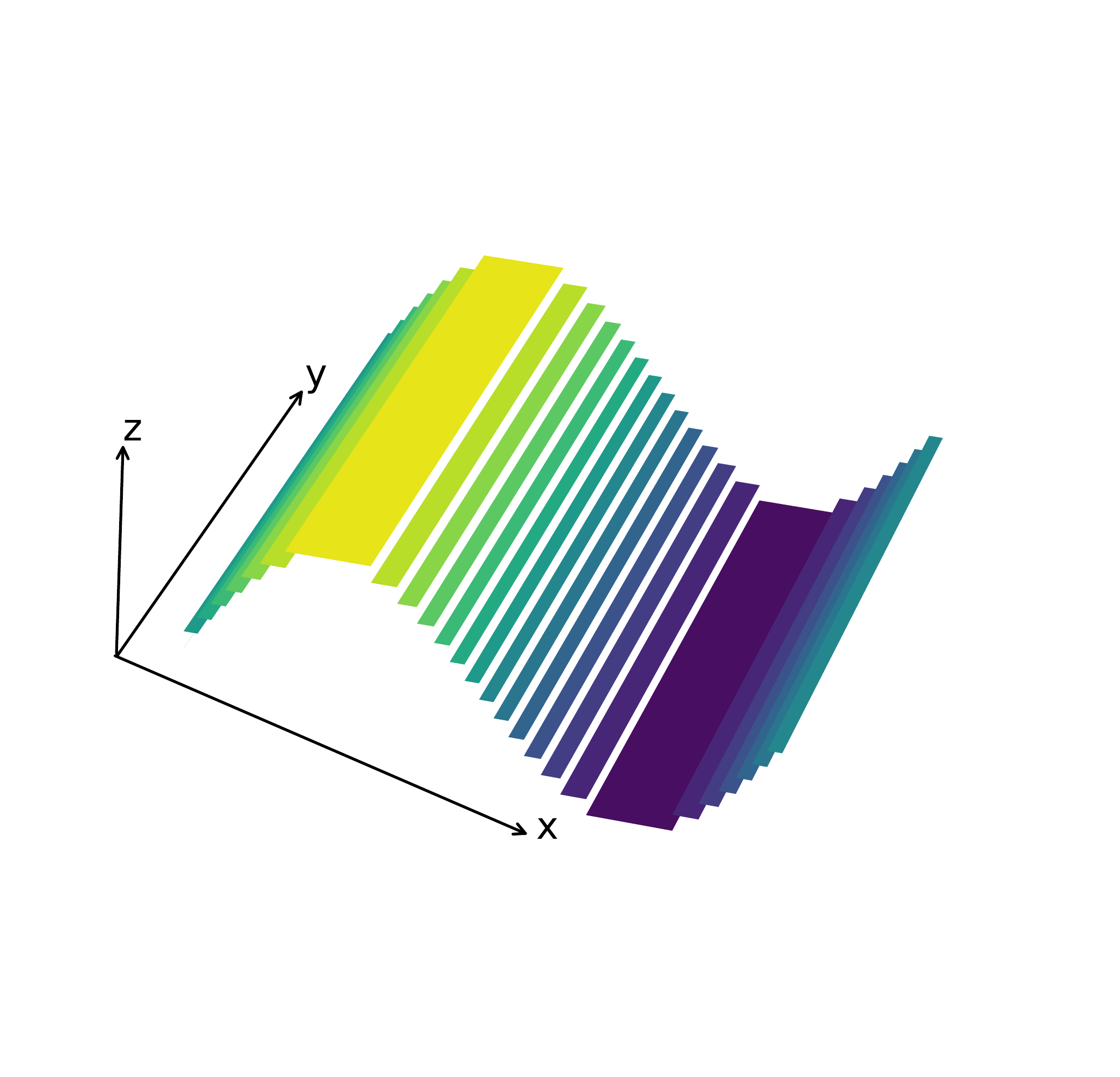}\label{f-deformation-a}}\quad
\subfloat[]{
\includegraphics[width=0.18 \textwidth]{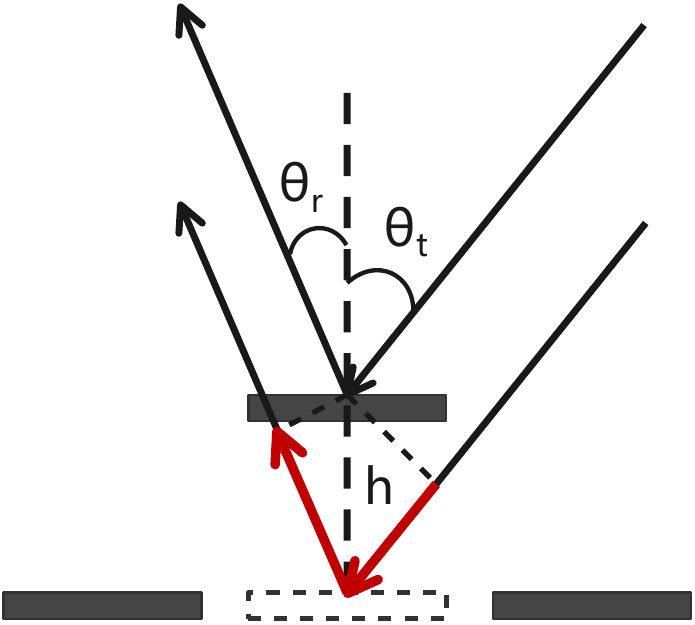}\label{f-deformation-b}}
\caption{(a) RIS panel with sine shaped deformation in x-direction; (b) path-difference due to element displacement in vertical direction. The coordinate in z-direction is scaled up in order to magnify the displacement in z-direction.}
\label{f-ris-deformation}
\end{figure}
As illustrated in Fig. \ref{f-ris-deformation}a,  a RIS panel is distorted into a sine-shaped surface due to interior stresses at x-direction. The deformation leads to different displacements to the RIS elements in z-direction which give rise to extra path difference to the incident and reflection wave between the adjacent elements (Fig. \ref{f-deformation-b}). Finally, the path differences among the RIS element give rise to a wave-front misalignment in the target direction and degrades the beamforming gain. In fact, there is also displacement in x-direction but is negligible compared with that in z-direction. The panel deformation caused by interior or exterior stresses is usually in regular shapes, such as sine-shaped, spherical or ellipsoidal surfaces (either concave or convex). When the RIS panel is formed by several sub-panels, V-shaped or step-shaped deformations may also occur when the panels are not in the same height in assembling. Nevertheless, the elements can be divided into groups with the elements in the same groups having the same displacement. 

\textbf{Remark 5. }If the incident and reflection beams are in the same plane, the phase-shift error of the $v$-th group elements due to panel deformation can be calculated as $\Delta\phi_v=kh(\cos\theta_i+\cos\theta_r)$ according to Fig. \ref{f-deformation-b}. Since the phase-shift error are fixed and identical within each group, then $\xi_v=1$, $\zeta_{vu}=\cos(\Delta\phi_v-\Delta\phi_u)$. Inserting these values to \eqref{eq-ep-2} yields
\begin{equation}
\delta=\frac{1}{N^2}\left[\sum_{v=1}^{V}N_{v}^2+2\sum_{v=1}^{V}\sum_{u=v+1}^{V}N_{v}N_{u}\cos(\Delta\phi_{v}-\Delta\phi_{u})\right].\label{eq-ep-deformation}
\end{equation}
The beamforming gain loss of a RIS with panel deformation in any shape can now be evaluated explicitly by applying \eqref{eq-ep-deformation}.

The sine-shaped deformation in Fig. \ref{f-ris-deformation}a is a quite special case in which the elements in the same column have the same displacement in z-direction and thus the RIS elements can be divided into $N_x$ groups with each group having $N_y$ elements. For this case, \eqref{eq-ep-deformation} can be simplified as
\begin{equation}
\delta=\frac{1}{N_x}+\frac{2}{N_x^2}\sum_{v=1}^{N_x}\sum_{u=v+1}^{N_x}\cos(\Delta\phi_{v}-\Delta\phi_{u}).\label{eq-ep-deformation-4}
\end{equation}
As long as the displacement of the element in each column can be determined, the beamforming gain loss can be estimated with \eqref{eq-ep-deformation-4}.

\begin{figure}[h]
\centering{\includegraphics[width=0.96 \textwidth]{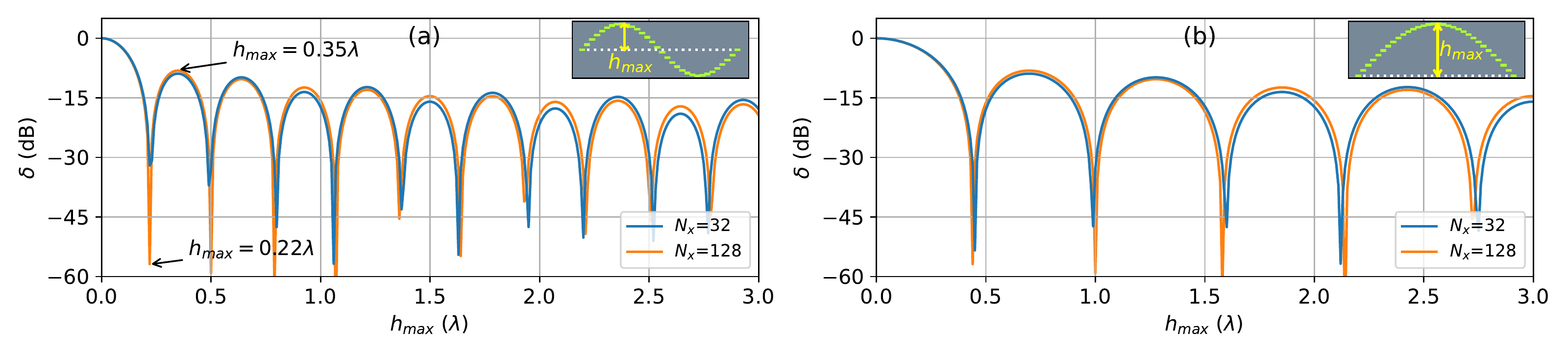}}
\caption{Theoretical gain loss versus maximum displacement $h_{\max}$ for (a) full 2 $\pi$ period sine-shaped deformation (Type A) and (b) half-period sine-shaped deformation (Type B) with $\theta_t=20^{\circ}$, $\varphi_t=0^{\circ}$ and $\theta_r=36^{\circ}$, $\varphi_r=180^{\circ}$.}
\label{f-sine-deformation}
\end{figure}
The theoretical bemforming gain losses of two typical types (Type A and B)  of deformations plotted in Fig. \ref{f-sine-deformation} show obvious periodicity versus the maximum element displacement $h_{\max}$ and the $2\pi$ period of the phase accounts for this periodicity. Consequently, a larger deformation does not necessarily result in a higher power gain loss. For example, a despairing gain loss is observed at $h_{\max}=0.22\lambda$ while it bounces back to about -8 dB when $h_{\max}$ is increased to 0.35$\lambda$ for Type A deformation in Fig. \ref{f-sine-deformation}a. However, the overall trend of the beamforming gain is going down with the increase of $h_{\max}$.
\begin{figure}[!h]
\centering{\includegraphics[width=0.98 \textwidth]{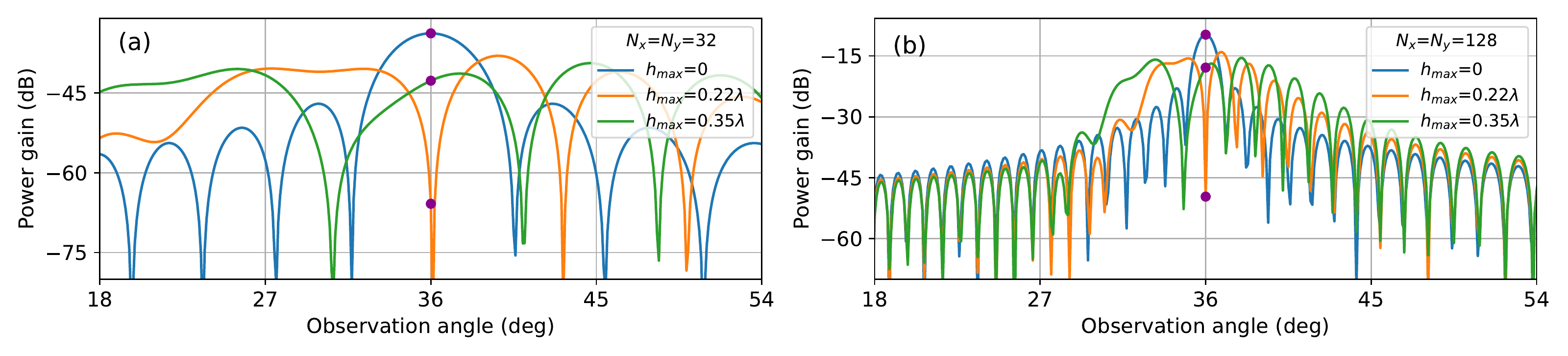}}
\caption{Simulation results of 2D beam patterns by RIS with sine-shaped deformation with different array sizes.}
\label{f-sine-deformation-2d}
\end{figure}

\begin{figure}[h]
\centering{\includegraphics[width=0.48 \textwidth]{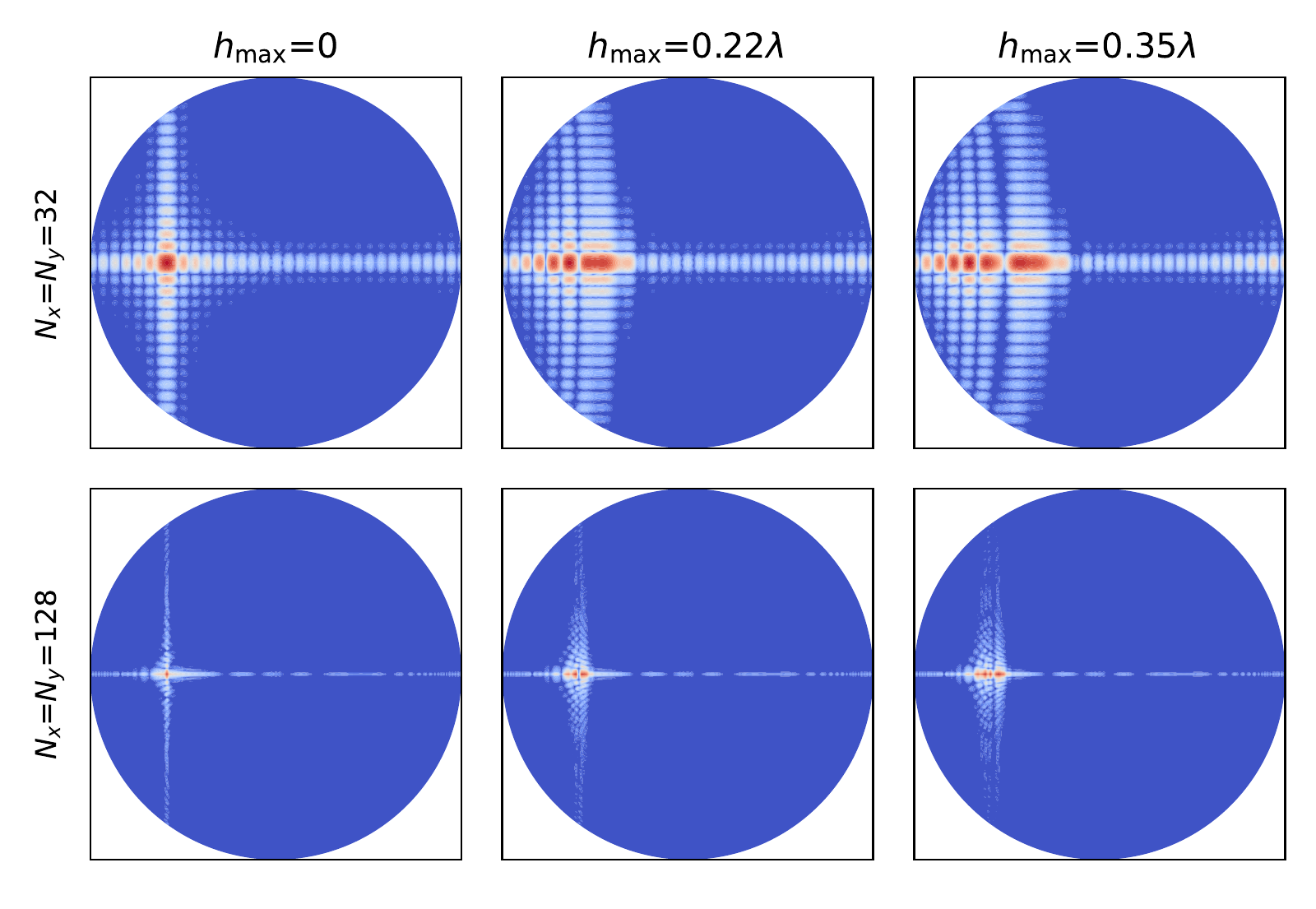}}
\caption{Power distribution over the sphere at $|\mathbf{r}|$=100 m. The upper and lower rows are simulated with 32$\times$32 and 128$\times$128 reflecting elements, respectively.}
\label{f-sine-deformation-3d}
\end{figure}

We are curious that how the reflected power has been redistributed under such RIS deformations and thus numerical simulations for Type A deformation are conducted and the results are shown in Fig. \ref{f-sine-deformation-2d} and \ref{f-sine-deformation-3d}. In Fig. \ref{f-sine-deformation-2d}, it can be found that RIS deformation may lead to a shift, spread or split to the main lobe. When $h_{\max}=0.22\lambda$, the main lobe splits right at the target direction ($\theta_r=36^{\circ}$) for $N_x=Ny=$128 while it split in close proximity to the target direction for $N_x=N_y=$32, and horrible gain drops are experienced. As to $h_{\max}=0.35\lambda$, however, the beam does not split in the vicinity of the target direction and the gain loss is relatively mild. The gain loss exceeds 3 dB when $h_{\max}>0.1\lambda$ for Type A deformation and $h_{\max}>0.2\lambda$ for Type B deformation, which sets quite a rigorous deformation limit for high-frequency RIS fabrication. For example, $h_{\max}$ should not be greater than 1 mm for a RIS at 30 GHz so that the power gain loss is less than 3 dB, which can be easily achieved in fabrication but may be hard to maintain after a long time of deployment. Fig. \ref{f-sine-deformation-3d}, on the other hand, indicates that larger $h_{\max}$ gives rise to severer beam spread and larger RIS has narrower beam spread as the displacement in z-direction between adjectent RIS elements are smaller for the same $h_{\max}$.

According to Fig. \ref{f-sine-deformation} - \ref{f-sine-deformation-3d}, it can be inferred that the impact of the RIS deformation on beamforming gain is more serious than the previously analyzed randomly distributed phase-shift errors. RIS deformations caused by environmental factors, such as temperature change, are common after deployment and these deformations may be visually tiny (for example, 1 mm) but is going to degrade the beamforming gain significantly.
\subsubsection{Wideband System Beamforming}
\begin{figure}[!h]
\centering
\subfloat[]{
\includegraphics[width=0.4 \textwidth]{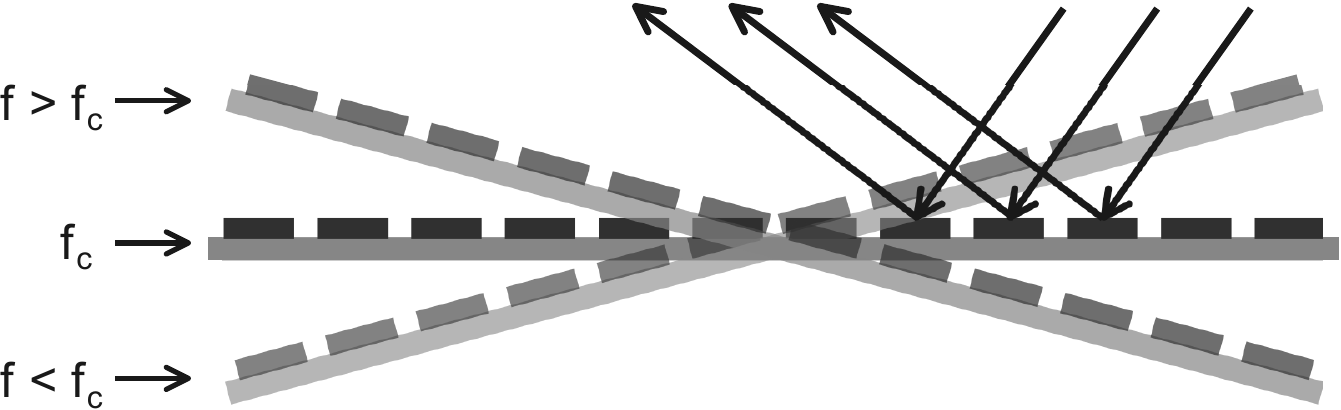}\label{f-wb-a}}
\subfloat[]{
\includegraphics[width=0.5 \textwidth]{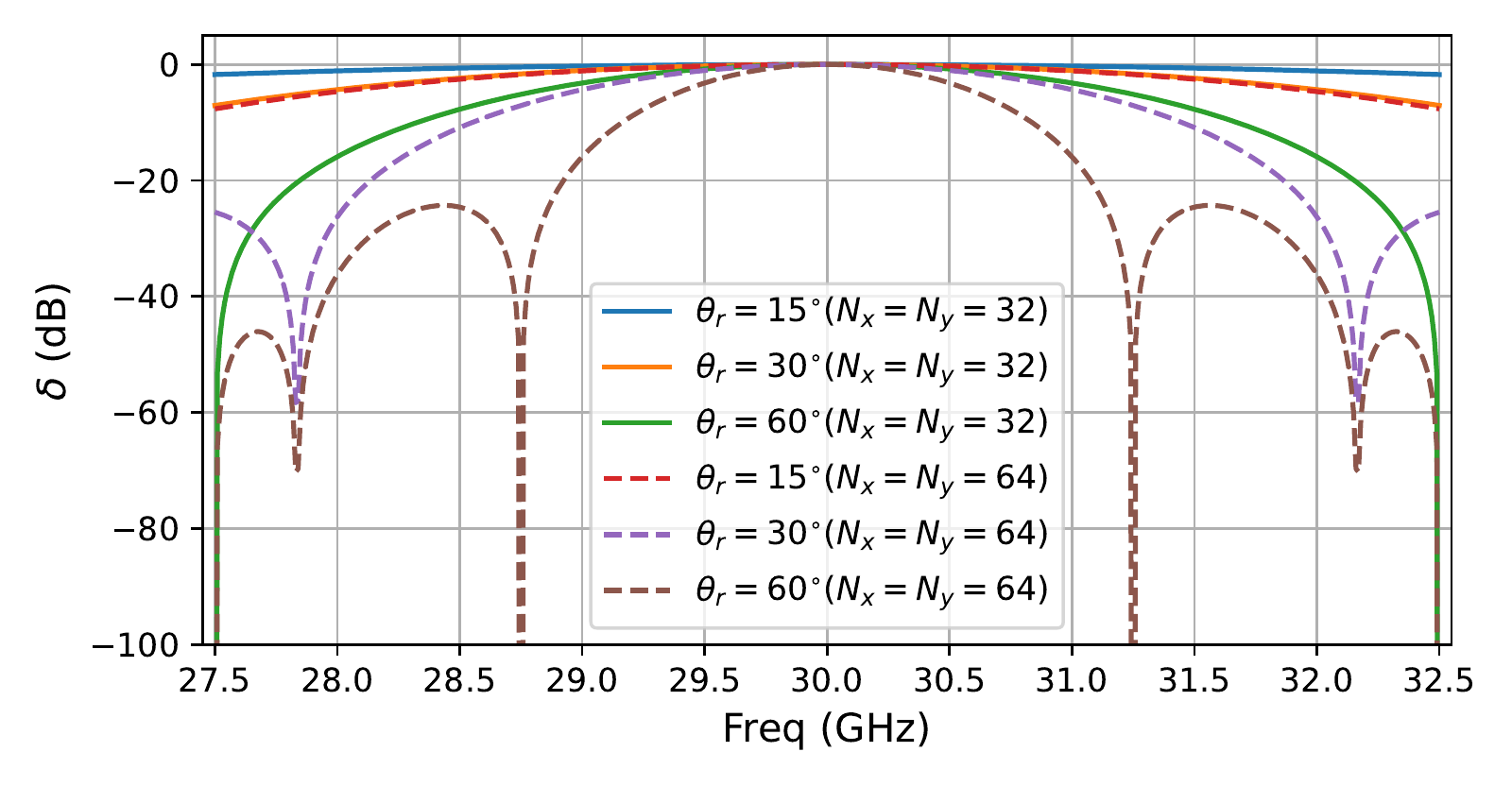}\label{f-wb-b}}
\caption{(a) Illustration of the impact of frequency mismatch in wideband system; (b) beamforming gain loss at different frequency when the precoding matrix is designed for the central frequency, 30 GHz. Normal incidence is assumed.}
\label{f-wb}
\end{figure}

Unlike the digital precoding at baseband, a RIS cannot provide individual phase-shift matrix for each sub-carrier and the phase-shifts for the RIS elements are usually designed for the central frequency $\nu_c$, i.e., $\phi_{\ell}^{*}=-k_{c}(|\mathbf{r}_t-\mathbf{r}_{\ell}|+|\mathbf{r}'-\mathbf{r}_{\ell}|)+\phi_c$ for the $\ell$-th element, but the ideal phase-shift for the $i$-th sub-carrier centered at $\nu_i$ should be $\phi_{\ell}^{*}=-k_{i}(|\mathbf{r}_t-\mathbf{r}_{\ell}|+|\mathbf{r}'-\mathbf{r}_{\ell}|)+\phi_c$, where $k_c=\frac{2\pi\nu_c}{c}$ and $k_i=\frac{2\pi\nu_i}{c}$ with  $c$ being the speed of light. For a wideband communication system, as a result, there are fixed phase-shift errors at the RIS elements among the sub-carriers away from the central frequency, $\Delta\phi_{\ell}=-\Delta k_i (|\mathbf{r}_t-\mathbf{r}_{\ell}|+|\mathbf{r}'-\mathbf{r}_{\ell}|)$, where $\Delta k_i=k_c-k_i$. Considering the far-field precoding \eqref{eq-phi-opt-ff}, the phase-shift error of the RIS element at the n-th row and m-th column for the $i$-th subcarrier becomes
\begin{equation}
\Delta\phi_{i,mn}=- \Delta k_i [(N_{y}-n)c_{y}+(N_{x}-m)c_{x}].\label{eq-dphi}
\end{equation}
There is a fixed phase-shift error increment between the adjacent elements along x- and y-direction which makes the frequency mismatch resemble a panel rotation or deformation as illustrated in Fig. \ref{f-wb}a. Such rotation or deformation does not reshape of the reflection beam but shift the beam direction.

\textbf{Remark 6. }Since the phase-shift error is fixed for each element, the beamforming gain of the i-th sub-carrier becomes
\begin{equation}
\begin{aligned}
P_{i} = &S|\alpha|^2\left|\sum_{m=1}^{N_x}\sum_{n=1}^{N_y}e^{j\Delta\phi_{i,mn}}\right|^2\\
=&S|\alpha|^2 \left|e^{-j\Delta k_{i}[(N_{x}-1)c_{x}+(N_{y}-1)c_{y}]}\frac{\left(1-e^{j\Delta k_{i}N_xc_x}\right)\left(1-e^{j\Delta k_{i}N_yc_y}\right)}{\left(1-e^{j\Delta k_{i}c_x}\right)\left(1-e^{j\Delta k_{i}c_y}\right)}\right|,
\end{aligned}\label{eq-sum-delta-phi}
\end{equation}
When $\nu_i=\nu_c$, $\Delta k_i=0$ and the fraction term in \eqref{eq-sum-delta-phi} has a limit $N_xN_y=N^2$. Finally, the beamforming gain loss for the i-th sub-carrier becomes
\begin{equation}
\delta_i=\frac{1}{N^2}\left|e^{-j\Delta k_{i}[(N_{x}-1)c_{x}+(N_{y}-1)c_{y}]}\frac{\left(1-e^{jN_xc_i}\right)\left(1-e^{jN_yd_i}\right)}{\left(1-e^{jc_i}\right)\left(1-e^{jd_i}\right)}\right|^2.\label{eq-freq-msimatch-gain}
\end{equation}
Let the frequency band of the wideband signal be $[\nu_1, \nu_2]$, then the total beamforming loss can be calculated as $\delta=\sum p_i\delta_i$, where $0\leq p_i\leq 1$ is the allocated power portion to the $i$-th subcarrier.

Obviously, the gain loss due to frequency mismatch in wideband system is not only relevant to the incident and reflection angles but also the number of RIS elements in x- and y-direction.  Fig. \ref{f-wb}b illustrates the relevance clearly: (i) the more the subcarrier deviates from the central frequency, the greater the loss, (ii) the larger the $\theta_r$, the greater the loss, and (iii) the larger the element array, the greater the loss. The first two observations can be explained as larger frequency deviation and larger reflection elevation angle lead to larger fixed phase-shift error according to \eqref{eq-dphi}, corresponding to larger rotation angle in Fig. \ref{f-wb}a, which makes the reflection beam further deviate from the target direction. As to the third observation, it is a consequence of the narrower beam with larger array.
\subsection{Summary}
\begin{table*}[b]
\caption{Phase-shift errors of RIS}
\label{t-summary}
\setlength{\tabcolsep}{2pt}
\renewcommand\arraystretch{1.2}
\resizebox{\textwidth}{!}{
\begin{tabular}{|c|c|c|c|c|c|c|}
\hline
Error type& \multicolumn{2}{c|}{Globally i.i.d errors} & \multicolumn{2}{c|}{Grouped i.i.d errors} & \multicolumn{2}{c|}{Grouped fixed errors}\\
\hline
Cause& \multicolumn{2}{c|}{quantization, fabrication errors} & \multicolumn{2}{c|}{fabrication errors, hardware failures} & \multicolumn{2}{c|}{\tabincell{c}{fabrication and assembling errors,\\ environmental factor}} \\
\hline
Model& \multicolumn{2}{c|}{$S\big|\sum_{\ell=1}^{N}\alpha_{\ell}e^{j\Delta\phi_{\ell}}\big|^2$} & \multicolumn{2}{c|}{$S\big|\sum_{v=1}^{V}\sum_{\ell=1}^{N_{v}}\alpha_{v\ell}e^{j\Delta\phi_{v\ell}}\big|^2$} & \multicolumn{2}{c|}{$S\big|\sum_{v=1}^{V}\sum_{\ell=1}^{N_{v}}\alpha_{v\ell}e^{j\Delta\phi_{v}}\big|^2$} \\
\hline
Typical distribution& \multicolumn{2}{c|}{$\mathcal{U}(\beta_{1},\beta_{2}),\mathcal{N}_{\text{t}}(\mu,\sigma,\mu-\psi,\mu+\psi)$} & \multicolumn{2}{c|}{$\mathcal{U}(\beta_{1},\beta_{2}),\mathcal{N}_{\text{t}}(\mu,\sigma,\mu-\psi,\mu+\psi)$} &\multicolumn{2}{c|}{$-$} \\
\hline
\multicolumn{1}{|c|}{Examples} & n-bit quantization & \tabincell{c}{n-bit quantization +\\$\mathcal{N}_t(0,\sigma,-\psi,\psi)$}  & \tabincell{c}{frequency mismatch\\(narrow band)} & PIN diode failures & sine-shaped deformation&\tabincell{c}{frequency mismatch\\(wide band)}\\
\hline
Gain loss& $\frac{2\left[1-\cos\left(\frac{2\pi}{2^{n+1}}\right)\right]}{\left(\frac{2\pi}{2^{n+1}}\right)^2}$ & \tabincell{c}{$e^{-\sigma^2}\frac{2\left[1-\cos\left(\frac{2\pi}{2^{n+1}}\right)\right]}{\left(\frac{2\pi}{2^{n+1}}\right)^2}$\\$\Big[\frac{\text{erf}\left(\frac{\psi-j\sigma^{2}}{\sqrt{2}\sigma}\right)+\text{erf}\left(\frac{\psi+j\sigma^{2}}{\sqrt{2}\sigma}\right)}{2 \text{erf}\left(\frac{\psi}{\sqrt{2}\sigma}\right)}\Big]^{2}$} & \tabincell{c}{$\frac{2}{\pi^2}\left[1-\cos(\Delta\psi)\right]$\\(1-bit)} & \tabincell{c}{$(1-p_1-p_2)^2$\\(1-bit)} & \tabincell{c}{$\frac{1}{N_x}+\frac{2}{N_x^2}\sum_{v=1}^{N_x}\sum_{u=v+1}^{N_x}$\\$\cos(\Delta\phi_{v}-\Delta\phi_{u})$}&\tabincell{c}{$\frac{1}{N^2}\Big|e^{-j\Delta k_{i}[(N_{x}-1)c_{x}+(N_{y}-1)c_{y}]}$\\$\frac{\left(1-e^{jN_xa_i}\right)\left(1-e^{jN_yb_i}\right)}{\left(1-e^{ja_i}\right)\left(1-e^{jb_i}\right)}\Big|^2$}\\
\hline
\end{tabular}
}
\end{table*}

The impacts of various phase-shift errors to the RIS beamforming gain have been analyzed quantitatively based on the beamforming model with phase-shift errors \eqref{eq-ep} in this section. Exact gain losses of several practical cases are derived and some of these cases are also verified with numerical results, which offer clearer views of the redistribution of the reflection energy under phase-shift errors. Finally, we summarize the above analysis with Table \ref{t-summary}.

The globally i.i.d error model was widely adopted in early studies and we extends the analysis to the hybrid phase-shift errors. The grouped i.i.d error model is a better model in practical RIS application as the RIS elements exhibit different phase-shift errors distribution at different state or in different working frequency. The grouped fixed phase-shift error model deals with another tough condition in RIS application that gives rise to deterministic phase-shift errors rather than randomly distributed errors.
\section{Conclusion}\label{sec-4}
A comprehensive investigation on the impact of phase-shift errors on the beamforming ability of RIS is presented in this paper. With an improved beamforming model, the impact of the phase-shift errors are analyzed under three categories: (1) globally i.i.d errors, (2) grouped i.i.d errors and (3) grouped fixed errors. Typical phase-shift errors, including quantization errors, frequency mismatch, PIN diode failure, fabrication errors and RIS panel deformation, are discussed. Although this paper focuses on the beamforming of the RIS, the phase-shift error model is also applicable to other passive or active phased arrays. With the theoretical and numerical analyses presented in Section \ref{sec-3}, the following conclusions can be drew.
\begin{itemize}
\item Randomly distributed phase-shift errors, whether globally or grouped i.i.d, have moderate impact on the RIS beamforming. For one thing, the randomness lowers the deviation of the phase-shift errors and mitigates the impact on the beamforming gain. For another, high precision assemble lines are able to confine the fabrication errors to an acceptable range.
\item If the phase-shift errors satisfy a random distribution that is composed of multiple uniformly distributed or truncated normally distributed variables, the total gain loss is the multiplication of the gain loss by each of these components. Hence, the power gain loss of a 1-bit RIS may be higher than the theoretical value, 3.9 dB, when there are fabrication errors. 
\item There will be gain loss when the RIS is not utilized at the designed frequency and the phase-shift error model can be applied for evaluation of the applicable frequency band  according to the phase response curve versus frequency at different states. The theoretical gain loss of the 1-bit RIS due to frequency mismatch indicates that the RIS may have a wider frequency band than we thought and the criteria for RIS design may be eased, but this is correct only when the RIS is applied for narrow band communications. 
\item PIN diode failures on a RIS also give rise to uniformly distributed phase-shift errors. When 30\% of the PIN diodes are out of order, a 3 dB loss is expected for a 1-bit RIS. Auto detection of the PIN diode failure for a commercial RIS production may be required so that the condition of the RIS can be evaluated on demand after deployment.
\item RIS panel deformation may be a big issue for RIS deployment as it may leads to significant beamforming gain loss, especially in high frequency band. Special backboards are in need for supporting the element array and to keep the flatness of the panel and special technique can be applied for releasing the accumulated interior stresses in the RIS panel.
\item RIS-assisted wideband communication is challenging due to the analog precoding. The quantitative analysis has shown severe beamforming gain loss, especially for the subcarriers on the fringe of the frequency band. New precoding algorithm or new RIS design paradigm may help solve the problem in wideband application.
\end{itemize}

{\appendix[Proof of the inner match factor $\xi_1$ for hybrid phase-shift errors]
For a group of RIS elements having i.i.d phase-shift errors which satisfy a specific random distribution $\Delta\phi\sim\mathcal{D}(\Delta\phi)$ with PDF $p(\Delta\phi)$ defined in $[a, b]$, it makes no difference to the RIS beamforming gain if the phase-shift errors is shifted by $-\frac{a+b}{2}$, i.e., $\Delta\phi^s=\Delta\phi-\frac{a+b}{2}$, according to \eqref{eq-eta-offset}. The inner match factor $\xi$ can then be calculated as
\begin{equation}
\xi=\mathbb{E}[\cos(\Delta\phi^s)]^2+\mathbb{E}[\sin(\Delta\phi^s)]^2,
\end{equation}
with
\begin{equation}
\left\{
\begin{aligned}
\mathbb{E}[\cos(\Delta\phi^s)]=&\int_{-t}^{t}p(\Delta\phi+\frac{a+b}{2})\cos\Delta\phi d\Delta\phi,\\
\mathbb{E}[\sin(\Delta\phi^s)]=&\int_{-t}^{t}p(\Delta\phi+\frac{a+b}{2})\sin\Delta\phi d\Delta\phi.
\end{aligned}\right.
\end{equation}
where $t=\frac{b-a}{2}$.
If $p(\Delta\phi)$ is symmetric about $\Delta\phi=\frac{a+b}{2}$, then $p(\Delta\phi+\frac{a+b}{2})$ is an even function over $(-t,t)$. Since sine function is odd, $\mathbb{E}[\sin\Delta\phi^s]=0$, which leads to $\xi=\mathbb{E}[\cos\Delta\phi^{s}]^2$.

Next, let us assume that,
\begin{itemize}
\item the phase-shift error $\Delta\phi$ consists of $n$ components, $\Delta\phi=\sum_{i=1}^{n}\Delta\phi^i$, where each of these components satisfies a random distribution $\Delta\phi^i\sim\mathcal{D}_i(\Delta\phi^i)$ with PDF $p_i(\Delta\phi^i)$ defined in $\Delta\phi^i\in(a_i, b_i)$,
\item $p_i(\Delta\phi^i)$ is symmetric about $\Delta\phi^i=\frac{a_i+b_i}{2}$ for $i=1,2,\cdots,n$.
\end{itemize}
We can perform the shift $\Delta\phi^{si}=\Delta\phi^i-\frac{a_i+b_i}{2}$ to each of the component so that the distributions of each error component are centered at $\Delta\phi=0$. By applying the trigonometric formula $\cos(x+y)=\cos x\cos y-\sin x \sin y$ to $\cos\sum_{i=1}^{n}\Delta\phi^{si}$ iteratively, we have
\begin{equation}
\cos\sum_{i=1}^{n}\Delta\phi^{si}=\prod_{i=1}^{n}\cos\Delta\phi^{si}+R,\label{eq-cos-decomp}
\end{equation}
where
\begin{equation}
R=\sum\left(\underbrace{\prod\cos\Delta\phi^{si}}_{n_1\; \text{terms}}\underbrace{\prod\sin\Delta\phi^{sj}}_{n_2\;\text{terms}}\right),i\neq j,n_1+n_2=n,
\end{equation}
is a summation of polynomials consisting of cosine and sine functions, each of which contains at least one sine function. Owing to the fact that sine function is odd while cosine function and the PDF $p_i$ are even over $(-\frac{b_i-a_i}{2},\frac{b_i-a_i}{2})$, we obtain $\mathbb{E}[R]=0$.
Similarly, $\sin\sum_{i=1}^{n}\Delta\phi^{si}$ can be expended by applying the trigonometric formula $\sin(x+y)=\sin x\cos y+\cos x\sin y$ iteratively and a summation of polynomials consisting of cosine and sine functions can be obtained and each of these polynomials contains at least one sine function. Consequently, we have $\mathbb{E}[\sin\sum_{i=1}^{n}\Delta\phi^{si}]=0$.
Finally, the inner match factor $\xi$ can be calculated as
\begin{equation}
\begin{aligned}
\xi=&\{\mathbb{E}[\cos\sum_{i=1}^{n}\Delta\phi^{si}]\}^2+\{\mathbb{E}[\sin\sum_{i=1}^{n}\Delta\phi^{si}]\}^2
=\prod_{i=1}^{n}\{\mathbb{E}[\cos\Delta\phi^{si}]\}^2=\prod_{i=1}^{n}\xi(\Delta\phi^i).
\end{aligned}
\end{equation}
}

\end{document}